# Magnetic Order and Symmetry in the 2D Semiconductor CrSBr

Kihong Lee,[1,†] Avalon H. Dismukes,[1] Evan J. Telford,[2] Ren A. Wiscons,[1] Xiaodong Xu,[3] Colin Nuckolls,[1] Cory R. Dean,[2] Xavier Roy,[1,†] Xiaoyang Zhu[1,†]

[1] Department of Chemistry, Columbia University, New York, NY 10027, USA

[2] Department of Physics, Columbia University, New York, NY 10027, USA

[3] Department of Physics and Department of Materials Science and Engineering, University of Washington, Seattle, WA 98195, USA

[†] To whom correspondence should be addressed.
XYZ (xyzhu@columbia.edu), XR (xr2114@columbia.edu), KL (kl2797@columbia.edu)

**The recent discovery of two-dimensional (2D) magnets**[1–3] **offers unique opportunities for the experimental exploration of low-dimensional magnetism**[4] **and the magnetic proximity effects**[5,6]**, and for the development of novel magnetoelectric, magnetooptic and spintronic devices**[7,8] **. These advancements call for 2D materials with diverse magnetic structures as well as effective probes for their magnetic symmetries, which is key to understanding intralayer magnetic order and interlayer magnetic coupling**[9–11]**. However, traditional techniques do not probe magnetic symmetry; these examples include magneto-optical Kerr effect**[2,3]**, reflective magnetic circular dichroism and Raman spectroscopy**[12–14]**, anomalous Hall effect**[15]**, tunneling magnetoresistance**[16,17]**, spin-polarized scanning tunneling microscopy**[9]**, and single-spin scanning magnetometry**[18]**. Here we apply second harmonic generation (SHG), a technique acutely sensitive to symmetry breaking, to probe the magnetic structure of a new 2D magnetic semiconductor, CrSBr. We find that CrSBr monolayers are ferromagnetically ordered below 146 K, an observation enabled by the discovery of a giant magnetic dipole SHG effect in the centrosymmetric 2D structure. In multilayers, the ferromagnetic monolayers are coupled antiferromagnetically, with the Néel temperature notably increasing with decreasing layer number. The magnetic structure of CrSBr, comprising spins co-aligned in-plane with rectangular unit cell, differs markedly from the prototypical 2D hexagonal magnets CrI$_3$ and Cr$_2$Ge$_2$Te$_6$ with out-of-plane moments.**

**Moreover, our SHG analysis suggests that the order parameters of the ferromagnetic monolayer and the antiferromagnetic bilayer are the magnetic dipole and the magnetic toroidal moments, respectively. These findings establish CrSBr as an exciting 2D magnetic semiconductor and SHG as a powerful tool to probe 2D magnetic symmetry, opening the door to the exploration of coupling between magnetic order and excitonic/electronic properties, as well as the magnetic toroidal moment, in a broad range of applications.**

SHG typically originates from the dominant electric dipole (ED) mechanism. ED SHG is forbidden in centrosymmetric materials, but it becomes nonzero when a magnetic phase transition breaks both space inversion and time reversal symmetries to produce time-noninvariant, or *c*-type, SHG, in contrast to the time-invariant *i*-type SHG[19–21]. While this symmetry breaking process was recently demonstrated in *c*-type SHG from the antiferromagnetic (AFM) CrI$_3$ bilayer[22], ED SHG cannot probe the ferromagnetic (FM) monolayer because the inversion symmetry persists across the magnetic phase transition. To fully characterize magnetic symmetry in centrosymmetric monolayers, we need to detect higher order contributions, particularly magnetic dipole (MD) SHG[23]. In this work, we demonstrate this capability by using SHG to probe the layer-dependent magnetic symmetry of the theoretically predicted 2D magnetic semiconductor CrSBr[24,25]. In the bulk, CrSBr is a layered van der Waals (vdW) A-type antiferromagnet with a bulk Néel temperature ($T_N$) of 132 K[26]. Each rectangular layer exhibits in-plane anisotropic FM order and these FM layers couple antiferromagnetically along the stacking direction[26]. In addition to its unique magnetic structure and high magnetic ordering temperature, CrSBr offers two additional distinguishing features compared to other 2D magnets: it is air stable and semiconducting as shown in recent transport measurements[27]. Electronic structure calculations predict that the CrSBr monolayer hosts a fully spin-polarized conduction band[24,25], opening the door to spintronic and magneto-opto-electronic applications.

Millimeter-size CrSBr single crystals were grown by chemical vapor transport from Cr and S$_2$Br$_2$[27,28]. Single crystal x-ray diffraction (SCXRD) confirms the layered vdW structure with orthorhombic *Pmmn* space group and structural anisotropy along all three lattice vectors (Fig. 1a, Table S1). CrSBr can be easily exfoliated to produce monolayer flakes with lateral sizes in the tens of μm range (Fig. 1b,c, Fig. S1). The thickness of an exfoliated monolayer measured by atomic force microscopy is 0.78 ± 0.03 nm (Fig. S2), in excellent agreement with the layer thickness determined from the bulk crystal structure (0.791 nm). The flat needle habit of the CrSBr crystals, a manifestation of the in-plane structural anisotropy, makes identification of the crystallographic directions easy; the long axis coincides with the crystallographic *a*-axis, as confirmed by SCXRD. Anisotropic peak responses in Raman spectra from bulk down to monolayer confirms that the anisotropic structural morphology persists when the crystals are exfoliated (Fig. 1d, also Fig. S3,4). Fig. 1e presents the room temperature photoluminescence (PL) and differential reflectance spectra of CrSBr flakes with 1-3 layer thickness (see Fig. S5 for PL from bulk crystal). With a PL peak at 1.28-1.29 eV and an absorption peak at 1.36-1.37 eV, the Stokes shifts of ~80 meV are within three times the thermal energy at room temperature, suggesting that PL originates from band-edge emission rather than localized ligand-field luminescence as seen in CrI$_3$[29]. These results are consistent with CrSBr being a semiconductor as suggested by theoretical calculations[24,25] and demonstrated in bulk transport measurements[27].

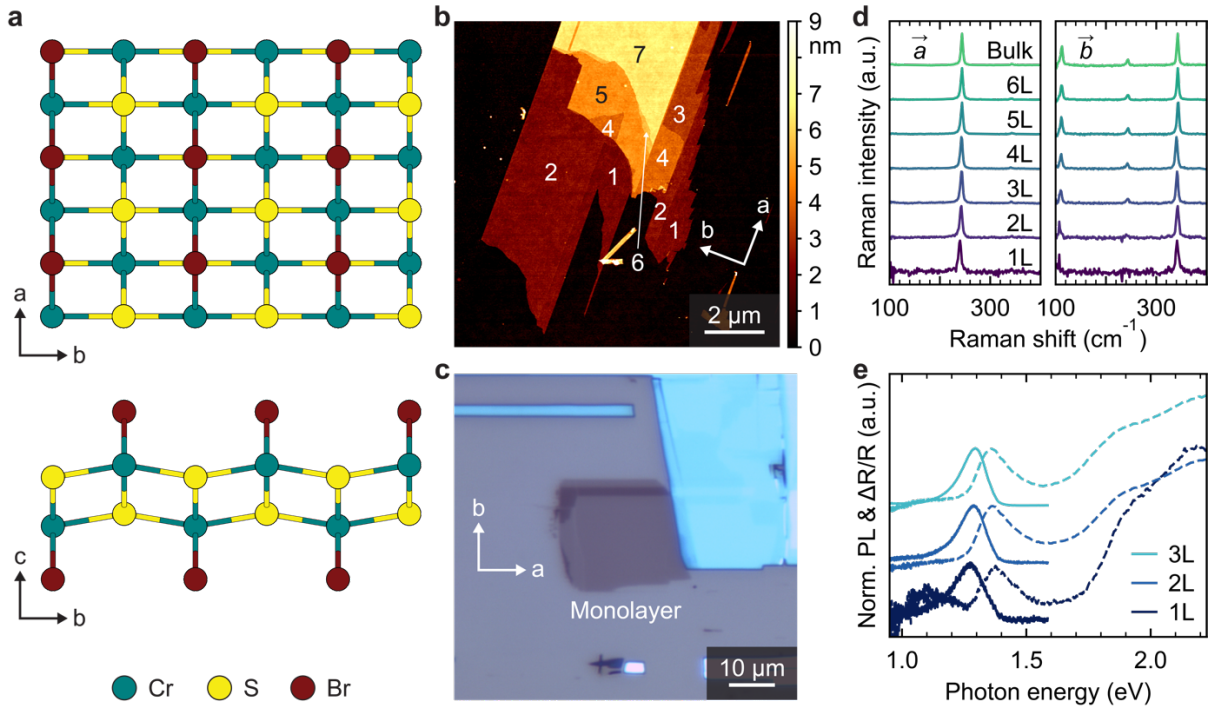

**Figure 1. a**, Crystal structure of CrSBr viewed along *c*- (top) and *a*-axes (bottom). **b**, Atomic force microscopy image of an exfoliated flake of varying thicknesses from 1 to 7 layers. **c**, Optical microscopy image of a CrSBr monolayer on a silicon substrate with 90 nm thermal oxide. **d**, Layer dependent Raman spectra (intensity normalized) of samples from one to six layers (1L-6L) and a thin bulk on a fused silica substrate with laser polarization along $\vec{a}$ and $\vec{b}$, respectively. **e**, Normalized photoluminescence (PL, solid) and differential reflectance (dashed) spectra at room temperature from one to three layers (1L-3L) of CrSBr on fused silica. The spectra are offset vertically for clarity.

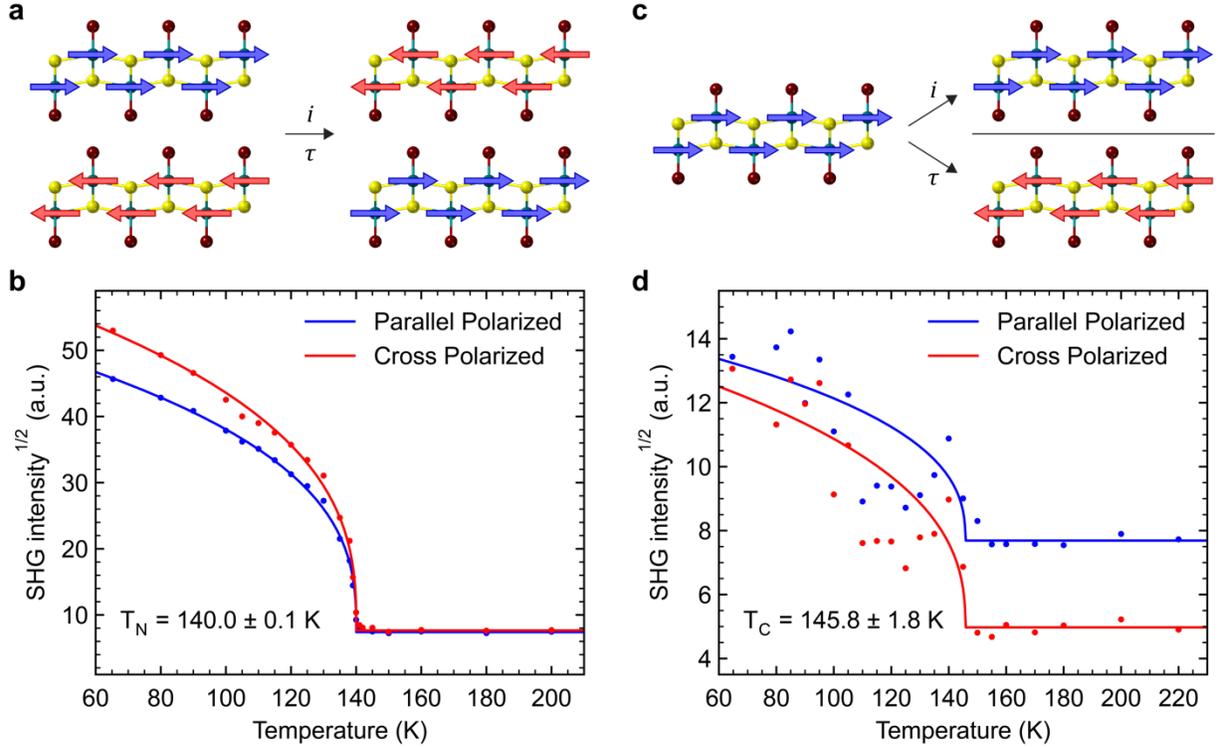

**Figure 2. a**, AFM order in CrSBr bilayer breaks inversion (*i*) and time reversal (*τ*) symmetries and allows ED SHG. **b**, Square roots of the average SHG intensity of bilayer (dots) as a function of temperature in CrSBr bilayer. The solid curves are fits to $(1 - T/T_c)^\beta$ to yield $T_N$ = 140.0 ± 0.1 K. **c**, The FM order of the CrSBr monolayer is centrosymmetric despite lacking time reversal symmetry. **d**, Square roots of the average SHG intensity of monolayer as a function of temperature with $T_C$ = 145.8 ± 1.8 K.

We probe magnetic order and phase transitions in CrSBr using SHG in parallel and cross polarization configurations, where fundamental and second harmonic polarizations are parallel and perpendicular to each other, respectively. We first focus on the CrSBr bilayer which, in its paramagnetic (PM) state, is centrosymmetric in the *mmm* ($D_{2h}$) point group (Fig. S6), and thus does not produce SHG intensity in the dominant ED mechanism. Below $T_N$, the interlayer AFM order breaks spatial inversion and time reversal symmetries (Fig. 2a) and the CrSBr bilayer should exhibit strong nonreciprocal, or time-noninvariant, SHG, similar to AFM $CrI_3$ bilayer[22]. Fig. 2b shows the temperature dependence of the square-root of SHG intensity, $I_{SHG}^{1/2} \propto |\chi^{(2),ED}|$, for both parallel and cross polarizations (PP, CP). Below 140 K, $I_{SHG}$ abruptly increases with decreasing temperature, signaling the loss of centrosymmetry as a result of the magnetic transition from the PM phase to the AFM phase. The solid curves are fits[30] to $|\chi^{(2),ED}| \propto M(T) = (1 - T/T_N)^\beta$, which give $T_N$ = 140.0 ± 0.1 K, higher than the bulk $T_N$ of 132 K. Furthermore, the fits yield β = 0.360 ± 0.006, indicating that the magnetic order of CrSBr bilayer follows the Heisenberg mechanism rather than the Ising (β ≈ 0.13) or the XY model (β ≈ 0.23)[26,31]. There is a weak and constant SHG background signal observed above $T_N$, which may originate from *i*-type interface or higher-order SHG contributions that are independent of the magnetic phase transition.

In contrast to the AFM CrSBr bilayer, the FM monolayer is centrosymmetric even with the broken time reversal symmetry (Fig. 2c) and does not produce SHG in the ED approximation. Unexpectedly, we again observe a rapid rise in SHG intensity below a critical temperature, signifying the phase transition from the PM to the FM state (Fig. 2d). Fits to $(1 - T/T_C)^\beta$ yield $T_C$ = 145.8 ± 1.8 K. Note that the SHG background signal above $T_C$ may have the same origin as that in Fig 2b.

To understand the origin of SHG from the centrosymmetric FM monolayer, we measured and modeled the polarization dependence of the SHG signals for both monolayer and bilayer CrSBr. In the AFM bilayer, the *c*-type ED SHG is symmetry allowed and the principal axis of rotation is a two-fold screw axis about the *a*-axis (Fig. 3a), which is also the direction of the magnetic toroidal moment (***T***, see below). Given the *mm*2 point group of the AFM bilayer, we calculate the second-order PP and CP susceptibilities $\chi_\parallel^{(2),ED}$ and $\chi_\perp^{(2),ED}$ following the procedure detailed in Tables S2-S4. The resulting fits describing the polarization dependent SHG intensity (solid lines; Fig. 3b) agree well with the SHG data measured in the AFM phase at 65 K (data point; Fig. 3b), and the principal axis of rotation determined from the model matches the crystallographic *a*-axis of the sample. As a comparison, the polarization-dependent SHG signal for the same sample in the PM phase at 200 K is nearly two-orders of magnitude weaker than that from the AFM bilayer (Fig. 3c). The angular distribution in the PM phase is more isotropic and is not described by the model.

The same cannot be said of the SHG response from the FM CrSBr monolayer, which stands in stark contrast with the SHG silent FM CrI$_3$ monolayer[22]. When the time reversal operator is considered, the magnetic point group of the FM CrSBr monolayer is *mmm* (classical subgroup 2/*m*), with the principal axis of rotation (two-fold screw axis) along the *b*-axis (Fig. 3d), which is also the direction of the magnetic moment (***M***). Given that the principal axis of rotation of the AFM bilayer is along the *a*-axis, the ED SHG polarization response of the FM monolayer should be rotated by π/2 from that of the AFM bilayer. Instead, we observe that the AFM bilayer and FM monolayer have very similar SHG polar profiles. Thus, the SHG response of the FM CrSBr monolayer does not originate from the ED mechanism, implying that the centrosymmetry is conserved across the magnetic transition, with no sign of internal canted magnetic structure that could destroy the inversion symmetry (Fig. S7b). We must consider SHG arising from higher order contributions[23]. While the ED SHG mechanism is symmetry forbidden, the axial tensors $\chi^{eem}$ and $\chi^{mee}$ governing the magnetic dipole contributions (eq. S8-S14 and associated text) are nonzero with inversion symmetry, making experimentally observable MD SHG a direct probe of the magnetization in the FM monolayer.

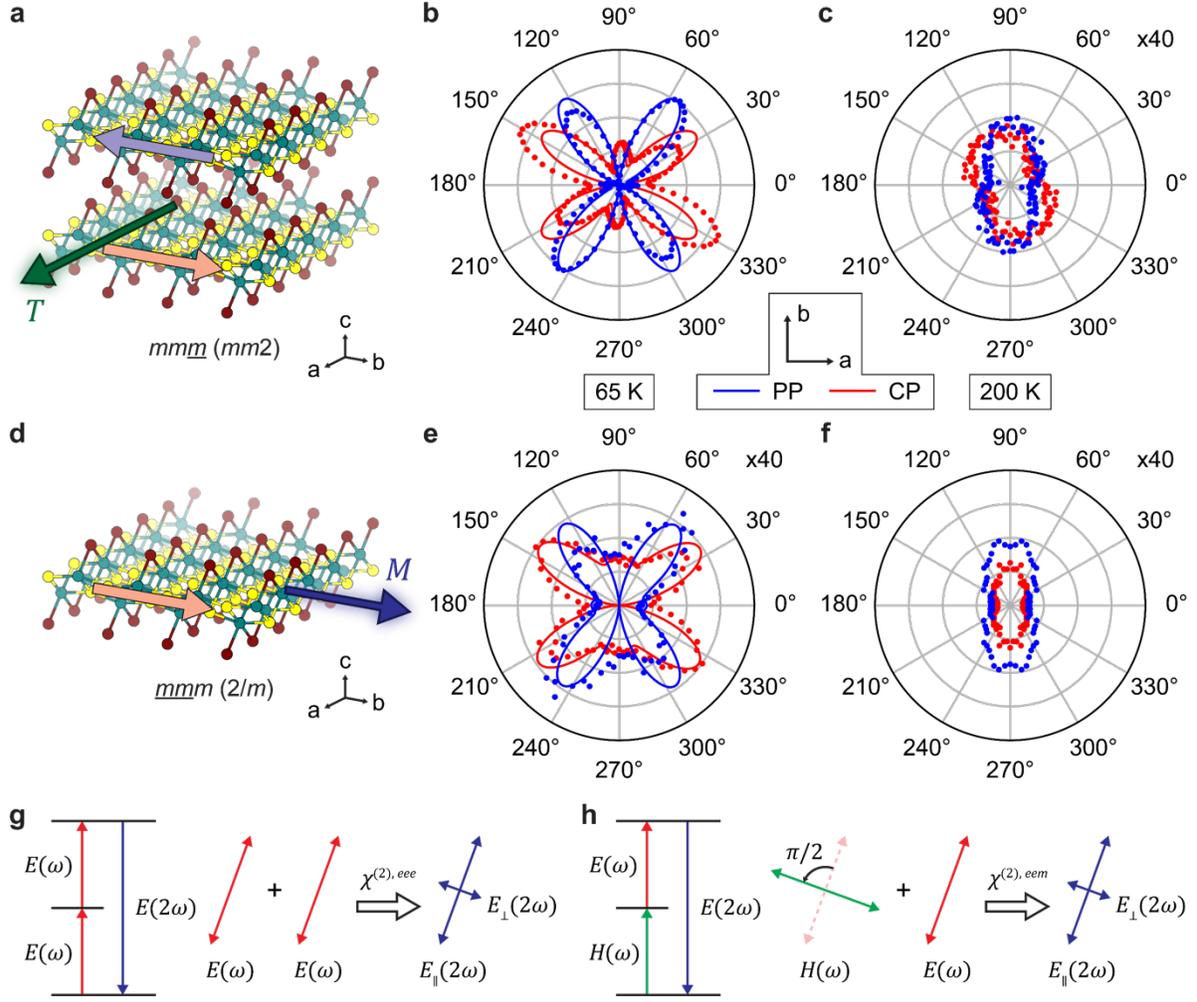

**Figure 3. a**, Magnetic symmetry of AFM bilayer, with its magnetic toroidal moment ***T*** as the order parameter. **b**, **c**, Polarization-resolved SHG (dots) from CrSBr bilayer in the AFM state at 65 K (**b**), and PM state at 200 K (**c**). Blue and red colors denote PP and CP excitation and detection configurations, respectively. The solid lines in (**b**) are fits to $|\chi^{(2),\text{ED}}|^2$ described by eq. S7. **d**, Magnetic symmetry of FM monolayer, with its net magnetization ***M*** as the order parameter. **e**, **f**, Polarization-resolved SHG (dots) from CrSBr monolayer in the FM state at 65 K (**e**), and PM state at 200 K (**f**). The solid lines in (**e**) are fits to the magnetic dipole $|\chi^{(2),\text{MD}}|^2$ described by eq. S13. Note that the intensities in (**c**), (**e**), and (**f**) are multiplied by a factor of 40, as compared to (**b**). **g**, **h**, Schematics of the electronic transitions and polarization rules for ED SHG (**g**) and MD SHG (**h**).

The MD susceptibilities $\chi_\parallel^{(2),MD}$ and $\chi_\perp^{(2),MD}$ calculated for the FM monolayer following the method described in eq. S8-S14 have the same functional forms as $\chi_\parallel^{(2),ED}$ and $\chi_\perp^{(2),ED}$ for the AFM bilayer, and the fits agree very well with the experimental data (Fig. 3e). As with the bilayer, the SHG spectra in the PM state is not described by the model, confirming that we are probing magnetic symmetry below $T_C$. Contributions from interface symmetry breaking and electric quadrupole could be comparable in magnitude to MD SHG, but they remain unchanged with or without time reversal symmetry and may only account for the background SHG signal above $T_C$. Figure 3g,h summarizes the electronic transitions and polarization rules for ED SHG and MD SHG and explains how two different mechanisms can produce the same polar profile. In essence, the change in the principal axis of rotation, from *a*-axis in the AFM bilayer to *b*-axis in the FM monolayer, is compensated by replacing the electric field polarization component of the electromagnetic wave with the perpendicular magnetic field polarization.

The MD SHG signal measured in the centrosymmetric FM CrSBr monolayer is much stronger than expected from the scaling relationship $|\chi^{(2),MD}|/|\chi^{(2),ED}| \sim a/\lambda \sim 10^{-3}$, where *a* is the unit cell size and $\lambda$ is the light wavelength[19]. This scaling law explains why MD SHG has been observed before in bulk magnetic solids such as NiO and $Cr_2O_3$[23,32], but not in 2D materials such as the $CrI_3$ monolayer[22]. To quantify the surprising MD SHG response of the FM CrSBr monolayer, we compare its magnitude to the SHG responses of the AFM CrSBr bilayer, AFM $CrI_3$ bilayer and $MoS_2$ monolayer. At 65 K with a 800 nm pump, the ED SHG intensity of the AFM CrSBr bilayer is ~10 times weaker than that of $MoS_2$ monolayer (Fig. S9) and comparable to that of AFM $CrI_3$ bilayer[22]. The MD SHG of the FM CrSBr monolayer is 50 times weaker than the ED SHG of AFM CrSBr bilayer. This translates to $|\chi^{(2),MD}|/|\chi^{(2),ED}| \sim 0.14$, about two orders of magnitude higher than expected from the $a/\lambda$ scaling[19]. The exact origin for this giant MD SHG effect from the FM CrSBr monolayer is unknown but we suspect it is related to the spin-orbit coupling in the intrinsic magnetic semiconductor.

Having established the MD SHG and nonreciprocal ED SHG mechanisms for the FM monolayer and AFM bilayer, respectively, we address the nature of the order parameter for each magnetic transition. For the FM monolayer, the order parameter is simply the net magnetization (***M***) along the easy *b*-axis (Fig. 3d). For the AFM bilayer, the order parameter cannot be the antiferromagnetic vector (i.e., the difference of the net magnetic vector of each layer) because it is noninvariant under symmetry operations of the magnetic order[30]. We instead identify the order parameter to be the magnetic toroidal moment given by $\boldsymbol{T} = \pm\frac{1}{2}dM\hat{a}$, where $d$ is the distance between two layers, $M$ is net magnetization of each layer, and $\hat{a}$ is the unit vector along $a$[33]. Despite that both $\chi^{(2),\mathrm{ED}}$ of AFM bilayer and $\chi^{(2),\mathrm{MD}}$ of monolayer inherit their values from *i*-type rank 4 axial tensors of *mmm* point group in their paramagnetic phases, they can possess different nonzero elements due to distinct orientations of their order parameters (see SI for details). The presence of magnetic toroidal moments in the AFM CrSBr bilayer offers the enticing possibilities of finding the magnetoelectric effect[34] and engineering the toroidal moments to enhance the SHG signal[35] by separating the two CrSBr layers with a 2D dielectric spacer.

We turn to the surprising finding that $T_N$ increases with decreasing layer number in CrSBr. In addition to AFM bilayer and FM monolayer (Fig. 2, and Fig. S10-S12), we measured SHG from a 6 layer CrSBr flake (Fig. S13) and determined its $T_N$ to be 137.9 ± 5.2 K (Fig. 4a). For comparison, $T_N$ = 132 K in bulk CrSBr (Fig. S14)[26,27]. Such an increase in the magnetic phase transition temperature contrasts with other 2D magnets[1–3] and seemingly contradicts conventional understanding, which predicts a decreased stability of the magnetically ordered phase as the materials approaches the 2D limit[8,36]. To explain this anomaly, we propose that CrSBr contains an intermediate magnetic phase (labeled iFM) above $T_N$ in which individual layers are ferromagnetically ordered internally but the interlayer coupling remains paramagnetic. A related spin polarized phase at 17.2 K has been observed for $CrCl_3$, which is slightly above the bulk AFM transition ($T_N$ = 14.1 K)[37–39]. The different magnetic phases of CrSBr and their proposed evolution as a function of layer number and temperature are illustrated in Fig. 4b,c. In bulk, the intralayer Curie temperature ($T_C^{\mathrm{intra}}$) is significantly higher than $T_N$ because the intralayer superexchange coupling is much stronger than the interlayer super-superexchange coupling (Fig. S15)[26]. Spin waves traveling along the *c*-direction can disturb the weak interlayer AFM coupling while maintaining the stronger intralayer FM order, resulting in the iFM phase. Decreasing the sample thickness confines and suppresses the spin wave excitation along the *c*-axis, further stabilizing the AFM phase and causing $T_N$ to increase toward $T_C^{intra}$ (Fig. 4c). The iFM phase is hidden in the SHG response because the centrosymmetry persists across the PM to iFM phase transition, explaining the apparent increase of $T_N$ with decreasing layer number. Concurrently, $T_C^{intra}$ decreases, following the universal behavior of a spin wave, which destabilizes magnetic order in low-dimensional systems.

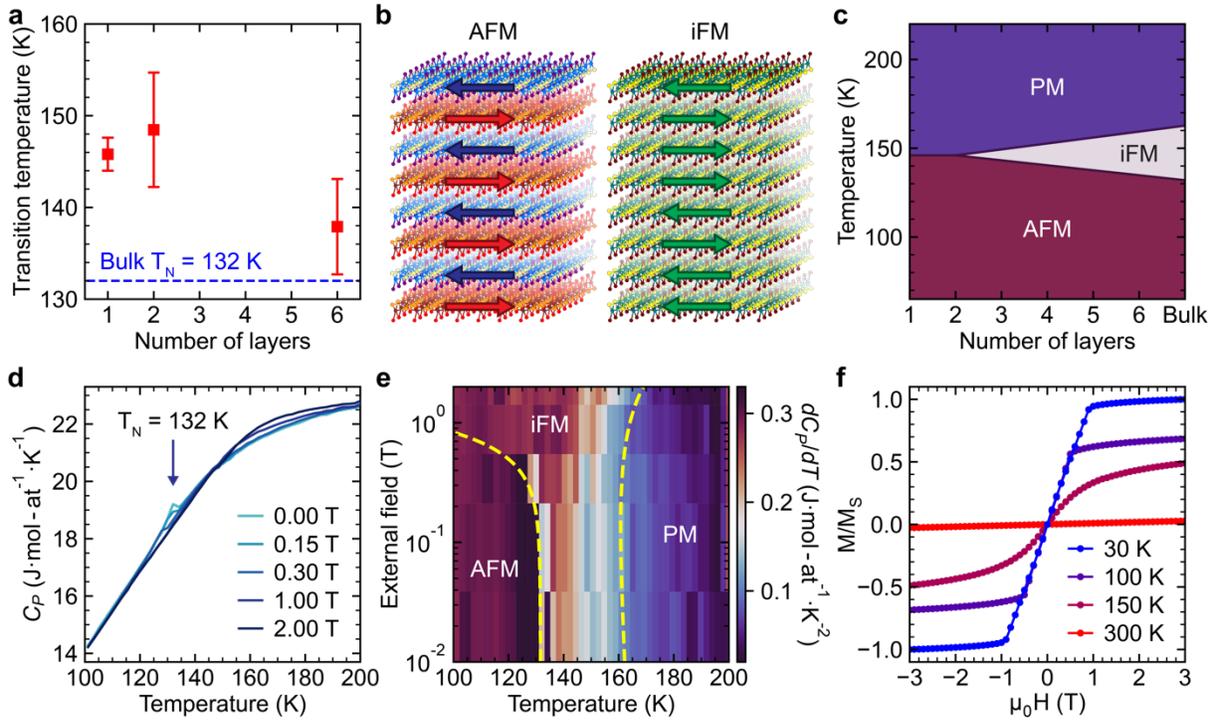

**Figure 4. a**, Magnetic phase transition temperature as a function of layer number, probed by SHG. **b**, Structures of AFM and iFM phases in bulk CrSBr at zero field. **c**, Schematic magnetic phase diagram presented as a function of layer number and temperature. **d**, Heat capacity of bulk CrSBr as a function of temperature at different applied magnetic field. The curves show a sharp AFM transition at 132 K and a broad feature around 160 K assigned to intralayer FM ordering. The external field is applied along the *c*-axis. **e**, Magnetic phase diagram of bulk CrSBr, as determined from $C_P$ measurements. The yellow dashed lines are guides to the eye to distinguish the different magnetic phases. **f**, Magnetization versus applied magnetic field at T = 30, 100, 150, and 300 K. M is normalized to the saturation magnetization value at T = 50 K. The magnetic field is applied along the *a*-axis. Data in (**e**) reproduced from ref. [27].

Bulk heat capacity ($C_P$) measurements support this magnetic phase diagram. The temperature dependence of $C_P$ at zero field displays a clear peak at 132 K corresponding to $T_N$ (Fig. 4d). The curve also shows a broad transition around 160 K (change in the slope) assigned to $T_C^{intra}$. The features assigned to $T_N$ and $T_C^{intra}$ correspondingly shift to lower and higher temperatures with increasing applied magnetic field, as expected for AFM and FM transitions, respectively. The phase diagram presented in Fig. 4e summarizes these results. In agreement with Fig. 4c, we note the decline of $T_C^{intra}$ with decreasing sample thickness, from ~160 K in the bulk to ~146 K in the monolayer. The iFM phase is also supported by isothermal field dependent magnetization (M) measurements (Fig. 4f, S16). Below $T_N$ (30 and 100 K), the antiferromagnetically coupled spins (along the magnetic easy *b*-axis) collectively and progressively align with the applied magnetic field (along the *a*-axis), until fully polarized above the saturation field. In the PM phase (300 K), M increases weakly and linearly with the applied field, never reaching saturation. The same PM behavior would be expected slightly above $T_N$ (150 K) but instead we measure a sigmoidal response evocating 2D superparamagnetic behavior. At 150 K, the material is in the iFM phase; it does not exhibit 3D long range magnetic order but the sigmoidal shape of M indicates that the 2D layers are ferromagnetically ordered internally.

Using second harmonic generation, we have demonstrated the emergence of ferromagnetic order in the 2D semiconductor CrSBr monolayer with a Curie temperature of $T_C$ = 146 ± 2 K. The identification of the FM order is enabled by a giant MD SHG effect with $\chi^{(2)}$ two orders of magnitude larger than expected. Multilayer CrSBr displays interlayer antiferromagnetic order and $T_N$ surprisingly increases with decreasing layer number, suggesting the presence of a hidden intralayer FM and interlayer PM phase between $T_N$ and $T_C$. In addition to magnetic symmetries, SHG analysis suggests magnetic dipole moment and magnetic toroidal moment as order parameters for the ferromagnetic monolayer and the antiferromagnetic bilayer, respectively. These findings establish the power of SHG in probing magnetic symmetry down to the 2D monolayer limit. Finally, the embodiment of both 2D magnetic and 2D semiconducting properties in a single material opens the door to exciting prospects of controlling one by the other. As the first step towards realizing this exciting potential, our preliminary results confirm that magnetic order in CrSBr is strongly coupled to optical transitions in temperature-dependent PL spectra, and to carriers in magnetotransport measurements showing large negative magnetoresistance (Supplementary Information II). Work is currently underway to understand the microscopic mechanisms of these couplings. Future research may explore the coupling between magnetic order and excitonic/electronic properties, as well as the magnetic toroidal moments, in a broad range of potential applications from coupled magnetic, spin, electronic, and optical processes.

**Acknowledgements**

The SHG measurements were supported by the US Air Force Office of Scientific Research (AFOSR) grant FA9550-18-1-0020 (to C.N., X.-Y.Z., and X. R.). Synthesis and structural characterization of CrSBr is supported by the Center on Programmable Quantum Materials, an Energy Frontier Research Center funded by the U.S. Department of Energy (DOE), Office of Science, Basic Energy Sciences (BES), under award DE-SC0019443. XYZ acknowledges partial support for laser equipment by the Vannevar Bush Faculty Fellowship through Office of Naval Research Grant # N00014-18-1-2080. C. N. thanks Sheldon and Dorothea Buckler for their generous support. A. H. D. is supported by the NSF graduate research fellowship program (DGE 16-44869). We thank Michael Spencer, Andrew Schlaus, Jue Wang, Yusong Bai for experimental assistance, Johannes Beck for helpful discussion on the synthesis, and Lucas Huber and Sebastian Maehrlein for discussions on SHG mechanisms.

**Author contributions.** XYZ, XR, and KL, conceived this work. KL performed all optical spectroscopic experiments. AHD grew the CrSBr crystals. KL, AHD, and EJT participated at various stages of sample preparation and characterization with supervisions from XR, XYZ, CN, and CRD. XYZ and XR supervised the project. XX participated in the interpretation of experimental findings. KL, XYZ, and XR wrote the manuscript, with inputs from all coauthors. All authors read and commented on the manuscript.

**Data Availability**. The source data presented in Figs. 1-4 are provided with the article. All other data that support the results in this article is available from the corresponding authors upon reasonable request.

**Competing Interests.** All authors declare that they have no competing interests.

**References**

1. Lee, J.-U. *et al.* Ising-type magnetic ordering in atomically thin FePS3. *Nano Lett.* **16,** 7433–7438 (2016).

2. Gong, C. *et al.* Discovery of intrinsic ferromagnetism in two-dimensional van der Waals crystals. *Nature* **546,** 265–269 (2017).

3. Huang, B. *et al.* Layer-dependent ferromagnetism in a van der Waals crystal down to the monolayer limit. *Nature* **546,** 270–273 (2017).

4. Mermin, N. D. & Wagner, H. Absence of ferromagnetism or antiferromagnetism in one-or two-dimensional isotropic Heisenberg models. *Phys. Rev. Lett.* **17,** 1133 (1966).

5. Zhong, D. *et al.* Layer-resolved magnetic proximity effect in van der Waals heterostructures. *Nat. Nanotechnol.* **15,** 187–192 (2020).

6. Tang, C., Zhang, Z., Lai, S., Tan, Q. & Gao, W. Magnetic Proximity Effect in Graphene/CrBr3 van der Waals Heterostructures. *Adv. Mater.* **32,** 1908498 (2020).

7. Gong, C. & Zhang, X. Two-dimensional magnetic crystals and emergent heterostructure devices. *Science* **363,** eaav4450 (2019).

8. Gibertini, M., Koperski, M., Morpurgo, A. F. & Novoselov, K. S. Magnetic 2D materials and heterostructures. *Nat. Nanotechnol.* **14,** 408–419 (2019).

9. Chen, W. *et al.* Direct observation of van der Waals stacking–dependent interlayer magnetism. *Science* **366,** 983–987 (2019).

10. Song, T. *et al.* Switching 2D magnetic states via pressure tuning of layer stacking. *Nat.*

*Mater.* **18,** 1298–1302 (2019).

11. McGuire, M. A., Dixit, H., Cooper, V. R. & Sales, B. C. Coupling of crystal structure and magnetism in the layered, ferromagnetic insulator CrI3. *Chem. Mater.* **27,** 612–620 (2015).

12. Huang, B. *et al.* Tuning inelastic light scattering via symmetry control in the two-dimensional magnet CrI 3. *Nat. Nanotechnol.* **15,** 212–217 (2020).

13. Jiang, S., Li, L., Wang, Z., Mak, K. F. & Shan, J. Controlling magnetism in 2D CrI3 by electrostatic doping. *Nat. Nanotechnol.* **13,** 549–553 (2018).

14. Huang, B. *et al.* Electrical control of 2D magnetism in bilayer CrI3. *Nat. Nanotechnol.* **13,** 544–548 (2018).

15. Liu, S. *et al.* Wafer-scale two-dimensional ferromagnetic Fe3GeTe2 thin films grown by molecular beam epitaxy. *npj 2D Mater. Appl.* **1,** 1–7 (2017).

16. Song, T. *et al.* Giant tunneling magnetoresistance in spin-filter van der Waals heterostructures. *Science* **360,** 1214–1218 (2018).

17. Klein, D. R. *et al.* Probing magnetism in 2D van der Waals crystalline insulators via electron tunneling. *Science* **360,** 1218–1222 (2018).

18. Thiel, L. *et al.* Probing magnetism in 2D materials at the nanoscale with single-spin microscopy. *Science* **364,** 973–976 (2019).

19. Fiebig, M., Pavlov, V. V & Pisarev, R. V. Second-harmonic generation as a tool for studying electronic and magnetic structures of crystals. *J. Opt. Soc. Am. B.* **22,** 96–118 (2005).

20. Kirilyuk, A. & Rasing, T. Magnetization-induced-second-harmonic generation from surfaces and interfaces. *JOSA B* **22,** 148–167 (2005).

21. Němec, P., Fiebig, M., Kampfrath, T. & Kimel, A. V. Antiferromagnetic opto-spintronics. *Nat. Phys.* **14,** 229–241 (2018).

22. Sun, Z. *et al.* Giant nonreciprocal second-harmonic generation from antiferromagnetic bilayer CrI3. *Nature* **572,** 497–501 (2019).

23. Fiebig, M. *et al.* Second harmonic generation in the centrosymmetric antiferromagnet NiO.

*Phys. Rev. Lett.* **87,** 137202 (2001).

24. Guo, Y., Zhang, Y., Yuan, S., Wang, B. & Wang, J. Chromium sulfide halide monolayers: intrinsic ferromagnetic semiconductors with large spin polarization and high carrier mobility. *Nanoscale* **10,** 18036–18042 (2018).

25. Wang, C. *et al.* A family of high-temperature ferromagnetic monolayers with locked spin-dichroism-mobility anisotropy: MnNX and CrCX (X= Cl, Br, I; C= S, Se, Te). *Sci. Bull.* **64,** 293–300 (2019).

26. Göser, O., Paul, W. & Kahle, H. G. Magnetic properties of CrSBr. *J. Magn. Magn. Mater.* **92,** 129–136 (1990).

27. Telford, E. J. *et al.* Layered Antiferromagnetism Induces Large Negative Magnetoresistance in the van der Waals Semiconductor CrSBr. *Arxiv Prepr.* **2005.06110,** (2020).

28. Beck, J. Über Chalkogenidhalogenide des Chroms Synthese, Kristallstruktur und Magnetismus von Chromsulfidbromid, CrSBr. *Zeitschrift für Anorg. und Allg. Chemie* **585,** 157–167 (1990).

29. Seyler, K. L. *et al.* Ligand-field helical luminescence in a 2D ferromagnetic insulator. *Nat. Phys.* **14,** 277–281 (2018).

30. Sa, D., Valenti, R. & Gros, C. A generalized Ginzburg-Landau approach to second harmonic generation. *Eur. Phys. J. B-Condensed Matter Complex Syst.* **14,** 301–305 (2000).

31. Bramwell, S. T. & Holdsworth, P. C. W. Universality in two-dimensional magnetic systems. *J. Appl. Phys.* **73,** 6096–6098 (1993).

32. Fiebig, M., Fröhlich, D., Krichevtsov, B. B. & Pisarev, R. V. Second harmonic generation and magnetic-dipole-electric-dipole interference in antiferromagnetic Cr2O3. *Phys. Rev. Lett.* **73,** 2127 (1994).

33. Dubovik, V. M. & Tugushev, V. V. Toroid moments in electrodynamics and solid-state physics. *Phys. Rep.* **187,** 145–202 (1990).

34. Spaldin, N. A., Fiebig, M. & Mostovoy, M. The toroidal moment in condensed-matter physics and its relation to the magnetoelectric effect. *J. Phys. Condens. Matter* **20,** 434203

(2008).

35. Van Aken, B. B., Rivera, J.-P., Schmid, H. & Fiebig, M. Observation of ferrotoroidic domains. *Nature* **449,** 702–705 (2007).

36. Zhang, R. & Willis, R. F. Thickness-dependent Curie temperatures of ultrathin magnetic films: effect of the range of spin-spin interactions. *Phys. Rev. Lett.* **86,** 2665 (2001).

37. Kuhlow, B. Magnetic ordering in CrCl3 at the phase transition. *Phys. status solidi* **72,** 161–168 (1982).

38. McGuire, M. A. *et al.* Magnetic behavior and spin-lattice coupling in cleavable van der Waals layered CrCl3 crystals. *Phys. Rev. Mater.* **1,** 14001 (2017).

39. Cai, X. *et al.* Atomically thin CrCl3: an in-plane layered antiferromagnetic insulator. *Nano Lett.* **19,** 3993–3998 (2019).

# Supporting information

**Magnetic Order and Symmetry in the 2D Semiconductor CrSBr**

Kihong Lee,[1,†] Avalon H. Dismukes,[1] Evan J. Telford,[2] Ren A. Wiscons,[1] Xiaodong Xu,[3] Colin Nuckolls,[1] Cory R. Dean,[2] Xavier Roy,[1,†] Xiaoyang Zhu[1,†]

[1] Department of Chemistry, Columbia University, New York, NY 10027, USA
[2] Department of Physics, Columbia University, New York, NY 10027, USA
[3] Department of Physics and Department of Materials Science and Engineering, University of Washington, Seattle, WA 98195, USA

[†] To whom correspondence should be addressed.
XYZ (xyzhu@columbia.edu), XR (xr2114@columbia.edu), KL (kl2797@columbia.edu)

**Table of Contents**

**Synthesis and structural characterization of bulk CrSBr**

Single crystals were synthesized using a modified procedure based on Beck's method (*1*). Disulfur dibromide ($S_2Br_2$) and elemental chromium in a 7:13 molar ratio were loaded into a quartz tube, which was then sealed under vacuum. The tube was heated in a temperature gradient of 950°C at the hot/reaction end and 850°C at the cold end. Crystals of the desired material were deposited in the center of the tube while $Cr_2S_3$ and $CrBr_3$ were found at the extremities. CrSBr crystals were cleaned by washing in pyridine, water, and then acetone.

      Single crystal X-ray diffraction (SCXRD) crystallography was performed using an Agilent SuperNova diffractometer with mirror-monochromated Cu Kα radiation. Data collection, integration, scaling (ABSPACK) and absorption correction (numeric analytical methods (*2*)) were performed in CrysAlisPro (*3*). Structure was solved using ShelXT (*4*), which was subsequent refined with full-matrix least-squares on F2 in ShelXL (*4*). Olex2 was used for viewing and preparing CIF files (*5*).

**Table S1.** Selected crystallographic data for CrSBr.

| | |
|---|---|
| *T (K)* | *100* |
| *Formula* | CrSBr |
| *MW* | 163.97 |
| *Space Group* | *Pmmn* |
| *a (Å)* | 4.7379 |
| *b (Å)* | 3.5043 |
| *c (Å)* | 7.9069 |
| *α (°)* | 90 |
| *β (°)* | 90 |
| *γ (°)* | 90 |
| *V (Å³)* | 131.28 |
| *Z* | 2 |
| *ρ$_{calc}$ (g cm$^{-3}$)* | 4.148 |
| *λ (Å)* | 0.71073 |
| *2θ$_{min}$, 2θ$_{max}$* | 10.032, 58.958 |
| *Nref* | 1685 |
| *R(int), R(s)* | 0.0735, 0.0406 |
| *μ (mm$^{-1}$)* | 19.976 |
| *Data* | 219 |
| *Restraints* | 0 |
| *Parameters* | 13 |
| *R$_1$ (obs)* | 0.0674 |
| *wR$_2$ (all)* | 0.1959 |
| *S* | 1.486 |

**Sample preparation**

CrSBr is micromechanically exfoliated to substrates using scotch tape. After cleaned with oxygen plasma, substrates are treated with 1-dodecanol to improve homogeneity of dielectric environment of silicon oxide surfaces (*6*). While arbitrary substrates can be chosen, silicon with 90 nm thermal oxide and UVFS are used for studies in this report. Flake thickness is identified with atomic force microscopy, which can be correlated to optical contrast for simplicity.

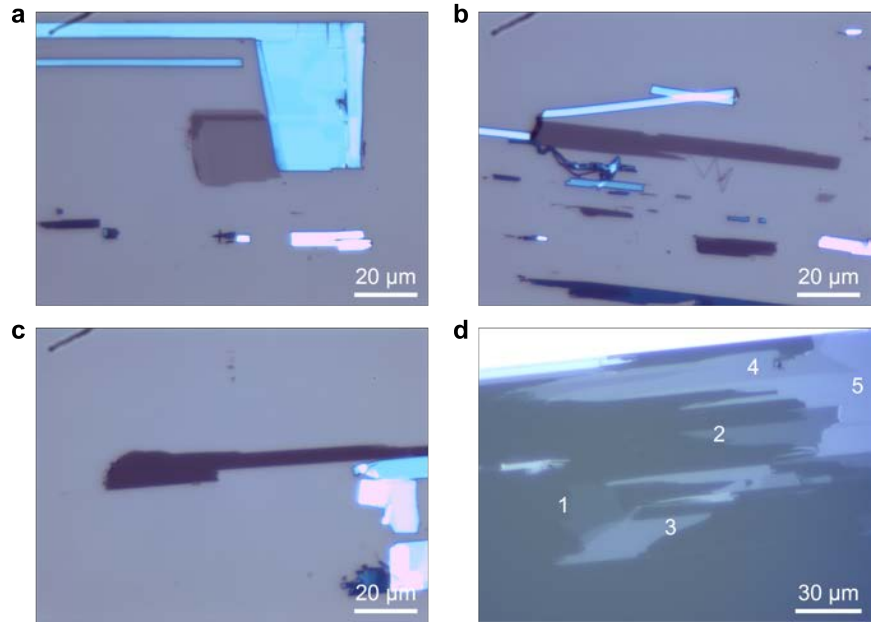

**Figure S1.** Optical microscope images of exfoliated CrSBr samples. **(a)** Monolayer (center piece) on a silicon substrate with 90 nm thermal oxide. **(b)** Bilayer (center piece) on a silicon substrate with 90 nm thermal oxide. **(c)** Trilayer on a silicon substrate with 90 nm thermal oxide. **(d)** 1-5 layers on UVFS.

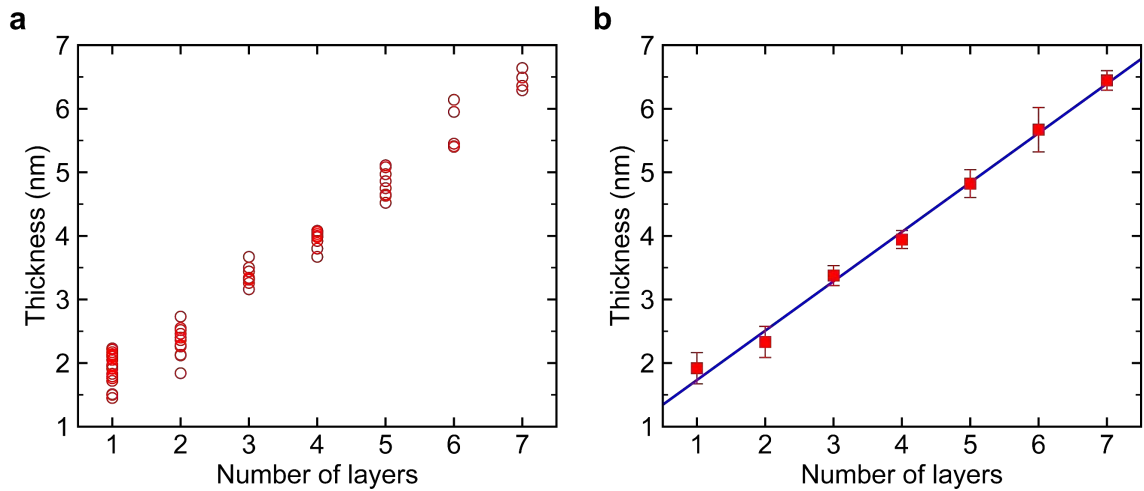

**Figure S2. (a)** Thicknesses versus the number of layers of various samples of exfoliated CrSBr. **(b)** Linear fit gives 0.78 ± 0.3 nm / layer, which agrees well with 0.791 nm obtained from single-crystal X-ray diffraction.

**Raman Spectroscopy**

Raman spectroscopy was performed with a Renishaw inVia confocal Raman microscope. 532 nm laser with ~1mW power was used for excitation. CrSBr flakes were exfoliated on 1-dodecanol passivated fused silica substrates, and samples were placed in a gas-tight cell with nitrogen atmosphere during the measurements. Excitation polarization was controlled with a half wave plate.

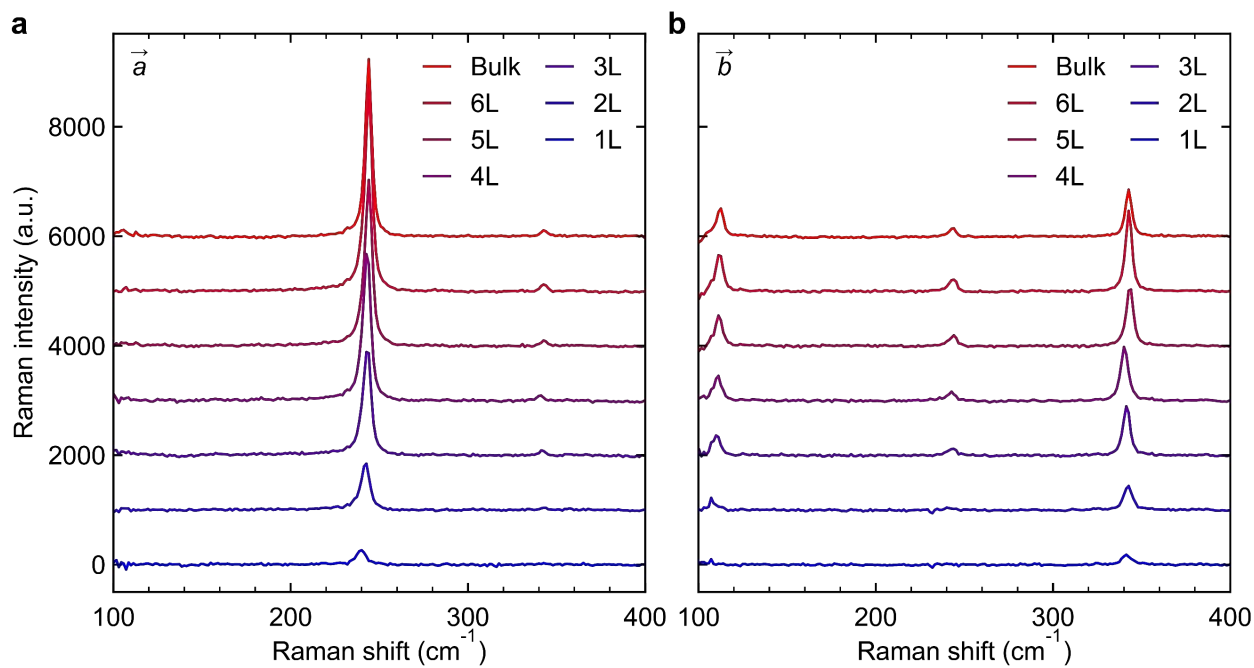

**Figure S3.** Raman spectra of samples from one to six layers and a thin bulk with excitation laser polarization along **(a)** $\vec{a}$ and **(b)** $\vec{b}$, respectively.

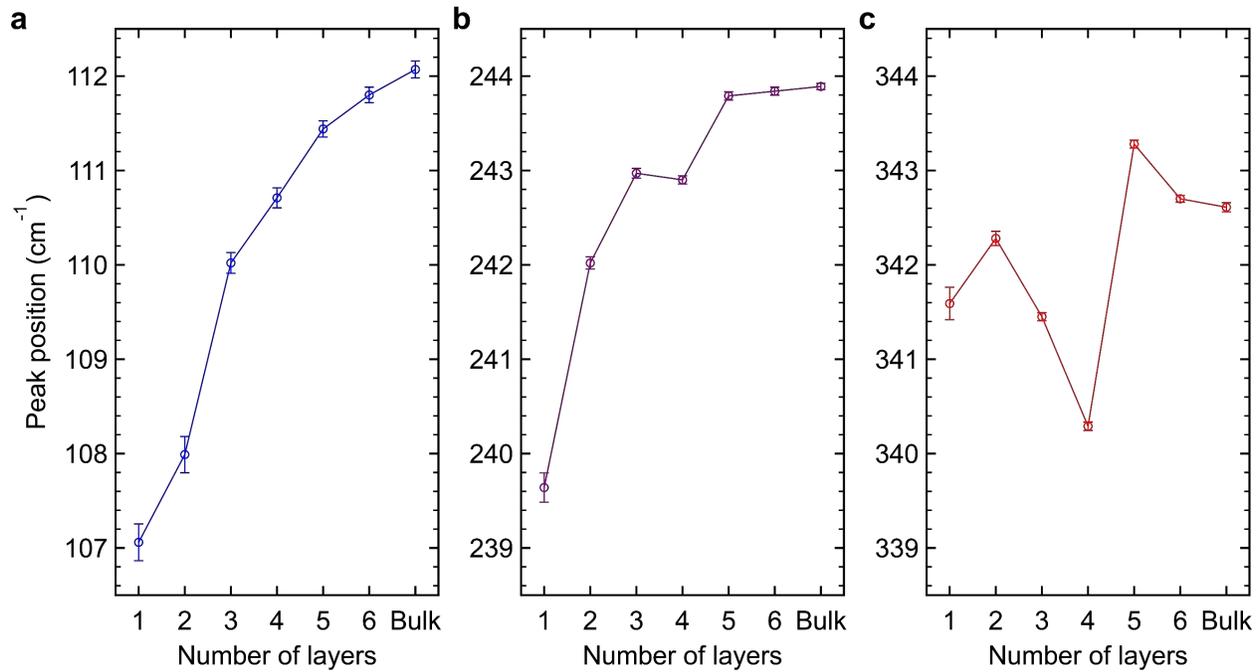

**Figure S4.** Raman peak positions as a function of layer number.

**Photoluminescence and Reflectance Spectroscopy**

Photoluminescence spectroscopy was performed with a 633 nm HeNe laser with on-sample power of 200 µW. Reflectance spectroscopy was conducted with broadband tungsten-halogen light source. Emission and reflection were collected with a Princeton Instruments PyLoN-IR and SpectraPro HRS-300. Samples were exfoliated on 1-dodecanol passivated UVFS, and were placed in nitrogen gas-tight cell throughout the measurements.

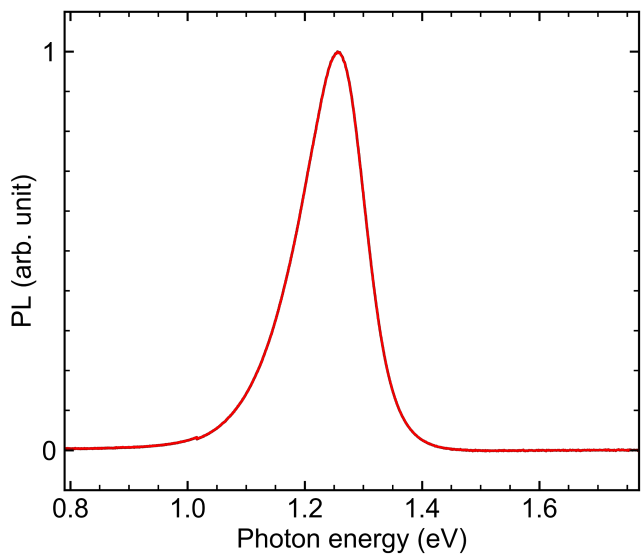

**Figure S5.** Photoluminescence spectrum of a bulk CrSBr crystal at room temperature.

**Magnetic symmetry of CrSBr bilayer and monolayer**

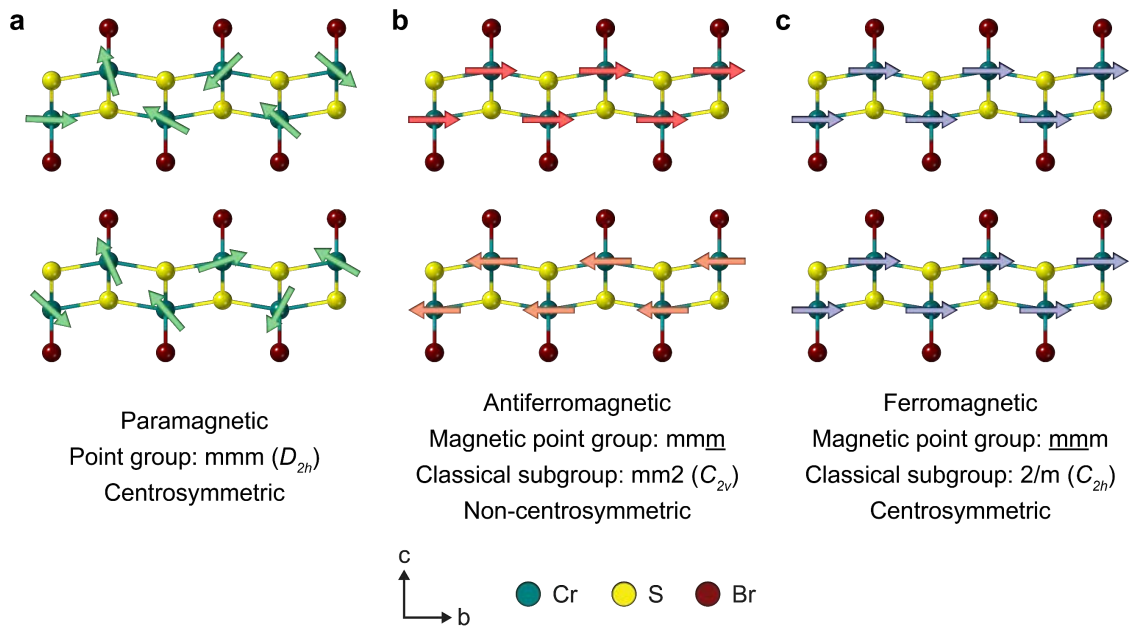

**Figure S6.** Magnetic point groups of **(a)** paramagnetic, **(b)** antiferromagnetic, and **(c)** ferromagnetic CrSBr bilayer. Only paramagnetic and antiferromagnetic states are experimentally observed.

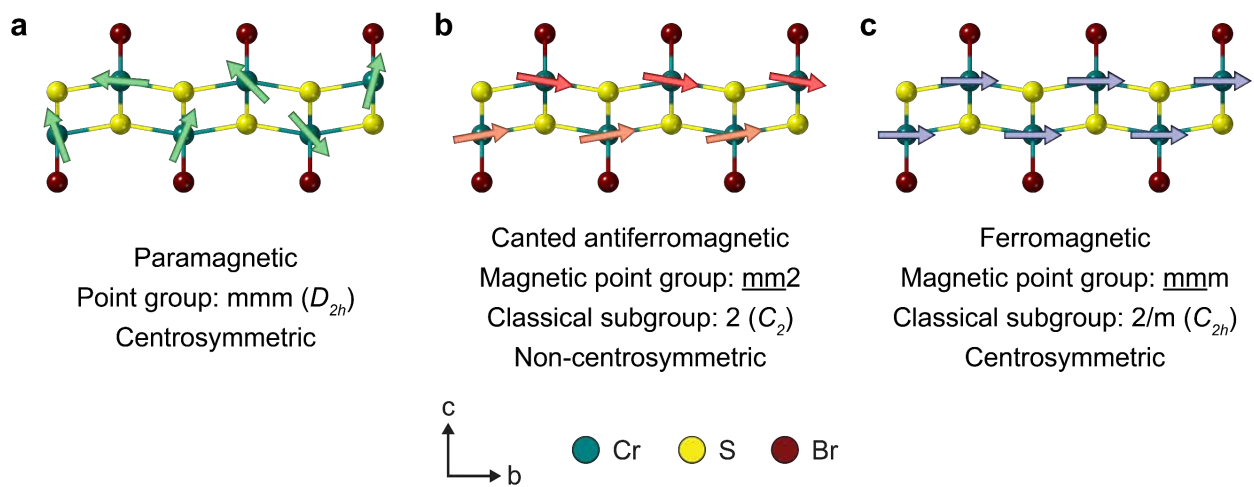

**Figure S7.** Magnetic point groups of **(a)** paramagnetic, **(b)** canted antiferromagnetic, and **(c)** ferromagnetic CrSBr monolayer. Only paramagnetic and ferromagnetic states are experimentally observed.

**Nonzero elements of second order susceptibilities**

**Table S2.** Nonzero elements of second order susceptibilities of *mmm*, *mmm̱*, and *m̱m̱m̱* point groups. Classical subgroups are given in parentheses.

| Point group | *mmm*, $D_{2h}$ | | *mmm̱* (*mm2*, $C_{2v}$) | | *m̱m̱m* (*2/m*, $C_{2h}$) | |
|---|---|---|---|---|---|---|
| Principal axis of rotation | *x*, *y*, or *z* | | *x* | | *y* | |
| Tensor type | Polar | Axial | Polar | Axial | Polar | Axial |
| Time-invariant (*i*-type) | 0 | $\chi^{(2)}_{xyz}, \chi^{(2)}_{xzy},$ $\chi^{(2)}_{yzx}, \chi^{(2)}_{yxz},$ $\chi^{(2)}_{zxy}, \chi^{(2)}_{zyx}$ | 0 | $\chi^{(2)}_{xyz}, \chi^{(2)}_{xzy},$ $\chi^{(2)}_{yzx}, \chi^{(2)}_{yxz},$ $\chi^{(2)}_{zxy}, \chi^{(2)}_{zyx}$ | 0 | $\chi^{(2)}_{xyz}, \chi^{(2)}_{xzy},$ $\chi^{(2)}_{yzx}, \chi^{(2)}_{yxz},$ $\chi^{(2)}_{zxy}, \chi^{(2)}_{zyx}$ |
| Time-noninvariant (*c*-type) | 0 | $\chi^{(2)}_{xyz}, \chi^{(2)}_{xzy},$ $\chi^{(2)}_{yzx}, \chi^{(2)}_{yxz},$ $\chi^{(2)}_{zxy}, \chi^{(2)}_{zyx}$ | $\chi^{(2)}_{xxx}, \chi^{(2)}_{xyy},$ $\chi^{(2)}_{xzz}, \chi^{(2)}_{yxy},$ $\chi^{(2)}_{yyx}, \chi^{(2)}_{zxz},$ $\chi^{(2)}_{zzx}$ | 0 | 0 | $\chi^{(2)}_{yyy}, \chi^{(2)}_{yxx},$ $\chi^{(2)}_{xyx}, \chi^{(2)}_{xxy},$ $\chi^{(2)}_{yzz}, \chi^{(2)}_{zyz},$ $\chi^{(2)}_{zzy}$ |

**Table S3.** Simplified table of nonzero elements of second order susceptibilities assuming back-scattering geometry and ignoring all terms containing *z* component.

| Point group | *mmm*, $D_{2h}$ | | *mmm̱* (*mm2*, $C_{2v}$) | | *m̱m̱m* (*2/m*, $C_{2h}$) | |
|---|---|---|---|---|---|---|
| Principal axis of rotation | *x*, *y*, or *z* | | *x* | | *y* | |
| Tensor type | Polar | Axial | Polar | Axial | Polar | Axial |
| Time-invariant (*i*-type) | 0 | 0 | 0 | 0 | 0 | 0 |
| Time-noninvariant (*c*-type) | 0 | 0 | $\chi^{(2)}_{xxx}, \chi^{(2)}_{xyy},$ $\chi^{(2)}_{yxy}, \chi^{(2)}_{yyx}$ | 0 | 0 | $\chi^{(2)}_{yyy}, \chi^{(2)}_{yxx},$ $\chi^{(2)}_{xyx}, \chi^{(2)}_{xxy}$ |

**Table S4.** Nonzero elements of second order susceptibilities of <u>mm</u>2 point group.

| Point group | <u>mm</u>2, (2, $C_2$) | | | <u>mm</u>2, (2, $C_2$) | |
|---|---|---|---|---|---|
| Principal axis of rotation | $y$ | | | $y$ | |
| Tensor type | Polar | Axial | | Polar | Axial |
| Time-invariant (*i*-type) | $\chi^{(2)}_{yyy}, \chi^{(2)}_{yxx},$ $\chi^{(2)}_{xyx}, \chi^{(2)}_{xxy},$ $\chi^{(2)}_{yzz}, \chi^{(2)}_{zyz},$ $\chi^{(2)}_{zzy}$ | $\chi^{(2)}_{xyz}, \chi^{(2)}_{xzy},$ $\chi^{(2)}_{yzx}, \chi^{(2)}_{yxz},$ $\chi^{(2)}_{zxy}, \chi^{(2)}_{zyx}$ | → Ignore elements with *z* component | $\chi^{(2)}_{yyy}, \chi^{(2)}_{yxx},$ $\chi^{(2)}_{xyx}, \chi^{(2)}_{xxy}$ | 0 |
| Time-noninvariant (*c*-type) | $\chi^{(2)}_{xyz}, \chi^{(2)}_{xzy},$ $\chi^{(2)}_{yzx}, \chi^{(2)}_{yxz},$ $\chi^{(2)}_{zxy}, \chi^{(2)}_{zyx}$ | $\chi^{(2)}_{yyy}, \chi^{(2)}_{yxx},$ $\chi^{(2)}_{xyx}, \chi^{(2)}_{xxy},$ $\chi^{(2)}_{yzz}, \chi^{(2)}_{zyz},$ $\chi^{(2)}_{zzy}$ | | 0 | $\chi^{(2)}_{yyy}, \chi^{(2)}_{yxx},$ $\chi^{(2)}_{xyx}, \chi^{(2)}_{xxy}$ |

**Second harmonic generation**

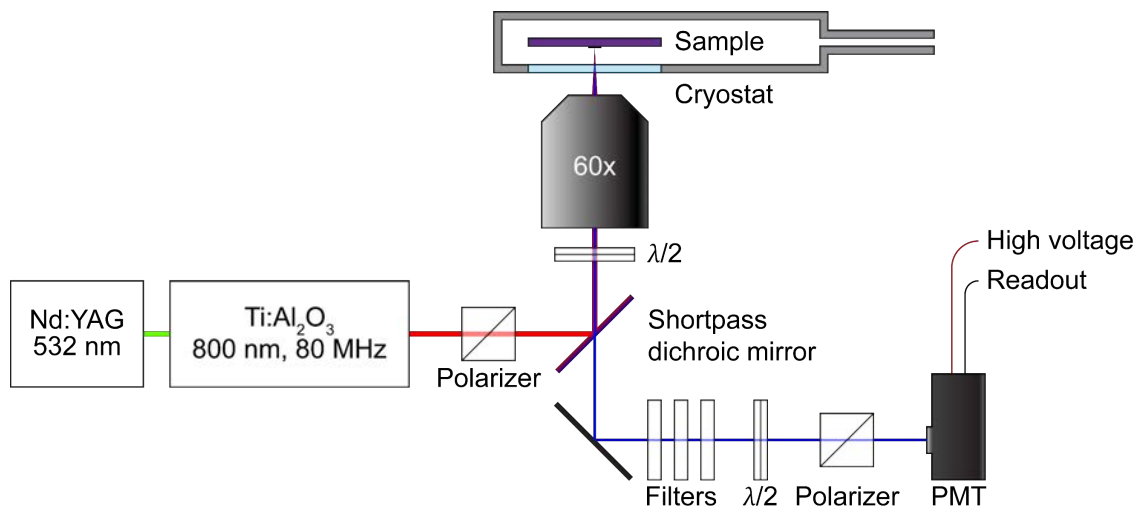

**Figure S8.** Schematics of SHG microscopy.

Second harmonic generation microscopy is performed with Spectra-Physics Tsunami 80 MHz Ti:Sapphire oscillator. 800 nm 420 µJ/cm$^2$ pulse with ~80 fs pulse-width is used to excite the sample. Achromatic half wave plate is placed between a shortpass dichroic mirror and a microscope objective to control polarizations of both incident pulse and emitting double-frequency radiation. A polarizer is placed before the dichroic to ensure polarization purity of the fundamental. Another half wave plate and a polarizer are placed before photomultiplier tube to choose between parallel and cross polarized configurations. Several filters are used to discriminate residual fundamental.

Sample is placed in an Oxford Instruments Microstat HiRes II with a 0.5 mm thick UVFS window. Base pressure is maintained at high vacuum around $10^{-7}$-$10^{-8}$ mbar. Sample is cooled with liquid nitrogen to reproducibly reach 65 K by pulling the cryogen with 0.1 atm. Liquid helium is used for some measurements to reach lower temperature.

Wave equation in a medium is described as

$$\vec{\nabla} \times (\vec{\nabla} \times \vec{E}) + \frac{1}{c^2} \frac{\partial^2 \vec{E}}{\partial t^2} = -\mu_0 \frac{\partial \vec{J}}{\partial t} \quad (S1)$$

where $\vec{J}$ is expanded into multipoles with

$$\vec{J} = \vec{J_0} + \frac{\partial \vec{P}}{\partial t} + \vec{\nabla} \times \vec{M} - \frac{\partial}{\partial t} \vec{\nabla} \cdot \hat{Q} + \cdots. \quad (S2)$$

Here we can define the source term

$$\vec{S} = \mu_0 \frac{\partial \vec{J}}{\partial t} = \mu_0 \frac{\partial^2 \vec{P}}{\partial t^2} + \mu_0 \vec{\nabla} \times \frac{\partial \vec{M}}{\partial t} + \cdots \quad (S3)$$

where nonlinear electric field follows $\vec{E}(2\omega) \propto \vec{S}^{(2)}$.

Electric dipole SHG is described by $P_i^{(2)} = \sum_{j,k} \varepsilon_0 \chi_{ijk}^{(2)} E_j E_k$ where $\vec{\chi}$ is a rank 3 tensor. Using permutation symmetry of $\chi_{ijk}^{(2)} = \chi_{ikj}^{(2)}$ induced second harmonic polarization can be written as

$$\begin{pmatrix} P_x(2\omega) \\ P_y(2\omega) \\ P_z(2\omega) \end{pmatrix} = \varepsilon_0 \begin{pmatrix} \chi_{xxx}^{(2)} & \chi_{xyy}^{(2)} & \chi_{xzz}^{(2)} & \chi_{xyz}^{(2)} & \chi_{xzx}^{(2)} & \chi_{xyx}^{(2)} \\ \chi_{yxx}^{(2)} & \chi_{yyy}^{(2)} & \chi_{yzz}^{(2)} & \chi_{yzy}^{(2)} & \chi_{yzx}^{(2)} & \chi_{yxy}^{(2)} \\ \chi_{zxx}^{(2)} & \chi_{yzz}^{(2)} & \chi_{zzz}^{(2)} & \chi_{zyz}^{(2)} & \chi_{zxz}^{(2)} & \chi_{zxy}^{(2)} \end{pmatrix} \begin{pmatrix} E_x(\omega)^2 \\ E_y(\omega)^2 \\ E_z(\omega)^2 \\ 2E_y(\omega)E_z(\omega) \\ 2E_z(\omega)E_x(\omega) \\ 2E_x(\omega)E_y(\omega) \end{pmatrix}. \quad (S4)$$

For *mmm* (classical subgroup *mm*2, or $C_{2v}$) point group of antiferromagnetic CrSBr bilayer, only *c*-type tensor is nonzero among the polar tensor. If we define the principal axis of rotation to be in *x*, the sensor is given as

$$\chi^{(2,c)} = \begin{pmatrix} \chi_{xxx}^{(2)} & \chi_{xyy}^{(2)} & \chi_{xzz}^{(2)} & 0 & 0 & 0 \\ 0 & 0 & 0 & 0 & 0 & \chi_{yxy}^{(2)} \\ 0 & 0 & 0 & 0 & \chi_{zxz}^{(2)} & 0 \end{pmatrix}. \quad (S5)$$

SHG microscopy is performed in back-scattering geometry where incident light and emitted light propagate in -*z* and *z* directions, respectively. In this configuration we can ignore contributions of $E_z(\omega)$ and $P_z(2\omega)$, which allows us to ignore any tensor element with *z* component and simplify the tensor to

$$\chi^{(2,c)} = \begin{pmatrix} \chi^{(2)}_{xxx} & \chi^{(2)}_{xyy} & 0 \\ 0 & 0 & \chi^{(2)}_{yxy} \end{pmatrix}. \tag{S6}$$

Assuming the incident light is linearly polarized whose polarization is rotated by an angle $\theta$ with respect to $x$ axis, parallel and cross polarized components of $\chi^{(2,c)}$ is written as

$$\begin{pmatrix} \chi^{(2)}_{\parallel} \\ \chi^{(2)}_{\perp} \end{pmatrix} = \begin{pmatrix} \cos\theta & \sin\theta \\ -\sin\theta & \cos\theta \end{pmatrix} \begin{pmatrix} \chi^{(2)}_{xxx} & \chi^{(2)}_{xyy} & 0 \\ 0 & 0 & \chi^{(2)}_{yxy} \end{pmatrix} \begin{pmatrix} \cos^2\theta \\ \sin^2\theta \\ 2\cos\theta\sin\theta \end{pmatrix}$$
$$= \begin{pmatrix} \chi^{(2)}_{xxx}\cos^3\theta + \left(\chi^{(2)}_{xyy} + 2\chi^{(2)}_{yxy}\right)\cos\theta\sin^2\theta \\ \left(-\chi^{(2)}_{xxx} + 2\chi^{(2)}_{yxy}\right)\cos^2\theta\sin\theta - \chi^{(2)}_{xyy}\sin^3\theta \end{pmatrix}. \tag{S7}$$

Since polar $\chi^{(2)}$ is zero at paramagnetic state with *mmm* point group, $\chi^{(2)}$ can probe magnetic phase transition of CrSBr bilayer

For magnetic dipole SHG of a centrosymmetric material, we consider both $\chi^{(2),eem}$ and $\chi^{(2),mee}$, whose SHGs are described with $P_i^{(2)} = \sum_{j,k} \varepsilon_0 \chi^{(2),eem}_{ijk} E_j H_k$ and $M_i^{(2)} = \sum_{j,k} \varepsilon_0 \chi^{(2),mee}_{ijk} E_j E_k$. For $\chi^{(2),eem}$ we cannot use the permutation symmetry for $j$ and $k$, and the equation becomes

$$\begin{pmatrix} P_x(2\omega) \\ P_y(2\omega) \\ P_z(2\omega) \end{pmatrix}$$
$$= \varepsilon_0 \begin{pmatrix} \chi^{(2)}_{xxx} & \chi^{(2)}_{xyy} & \chi^{(2)}_{xzz} & \chi^{(2)}_{xyz} & \chi^{(2)}_{xzy} & \chi^{(2)}_{xzx} & \chi^{(2)}_{xxz} & \chi^{(2)}_{xxy} & \chi^{(2)}_{xyx} \\ \chi^{(2)}_{yxx} & \chi^{(2)}_{yyy} & \chi^{(2)}_{yzz} & \chi^{(2)}_{yyz} & \chi^{(2)}_{yzy} & \chi^{(2)}_{yzx} & \chi^{(2)}_{yxz} & \chi^{(2)}_{yxy} & \chi^{(2)}_{yyx} \\ \chi^{(2)}_{zxx} & \chi^{(2)}_{yzz} & \chi^{(2)}_{zzz} & \chi^{(2)}_{zyz} & \chi^{(2)}_{zzy} & \chi^{(2)}_{zzx} & \chi^{(2)}_{zxz} & \chi^{(2)}_{zxy} & \chi^{(2)}_{zyx} \end{pmatrix} \begin{pmatrix} E_x(\omega)^2 \\ E_y(\omega)^2 \\ E_z(\omega)^2 \\ E_y(\omega)H_z(\omega) \\ E_z(\omega)H_y(\omega) \\ E_z(\omega)H_x(\omega) \\ E_x(\omega)H_z(\omega) \\ E_x(\omega)H_y(\omega) \\ E_y(\omega)H_x(\omega) \end{pmatrix},$$
$$\tag{S8}$$

and for $\chi^{(2),mee}$,

$$\begin{pmatrix} M_x(2\omega) \\ M_y(2\omega) \\ M_z(2\omega) \end{pmatrix} = \varepsilon_0 \frac{c}{n_{2\omega}} \begin{pmatrix} \chi^{(2)}_{xxx} & \chi^{(2)}_{xyy} & \chi^{(2)}_{xzz} & \chi^{(2)}_{xyz} & \chi^{(2)}_{xzx} & \chi^{(2)}_{xyx} \\ \chi^{(2)}_{yxx} & \chi^{(2)}_{yyy} & \chi^{(2)}_{yzz} & \chi^{(2)}_{yzy} & \chi^{(2)}_{yzx} & \chi^{(2)}_{yxy} \\ \chi^{(2)}_{zxx} & \chi^{(2)}_{yzz} & \chi^{(2)}_{zzz} & \chi^{(2)}_{zyz} & \chi^{(2)}_{zxz} & \chi^{(2)}_{zxy} \end{pmatrix} \begin{pmatrix} E_x(\omega)^2 \\ E_y(\omega)^2 \\ E_z(\omega)^2 \\ 2E_y(\omega)E_z(\omega) \\ 2E_z(\omega)E_x(\omega) \\ 2E_x(\omega)E_y(\omega) \end{pmatrix} \quad (S9)$$

where $c$ is the speed of light and $n_{2\omega}$ is the material index of refraction at frequency $\omega$.

Our material of interest is CrSBr monolayer, whose point group is *mmm* at its paramagnetic state and *mmm* at its ferromagnetic state. Since the both point groups are centrosymmetric, all polar tensors are zeros, and terms like $\chi^{(2),eee}$, $\chi^{(2),mem}$, $\chi^{(2),emm}$, and $\chi^{(3),eeem}$ can be ignored. For *mmm* group, nonzero elements for both *i*- and *c*-type axial tensors are $\chi^{(2)}_{xyz}$, $\chi^{(2)}_{xzy}$, $\chi^{(2)}_{yzx}$, $\chi^{(2)}_{yxz}$, $\chi^{(2)}_{zxy}$, and $\chi^{(2)}_{zyx}$. For *mmm* group, while *i*-type axial tensor has $\chi^{(2)}_{xyz}$, $\chi^{(2)}_{xzy}$, $\chi^{(2)}_{yzx}$, $\chi^{(2)}_{yxz}$, $\chi^{(2)}_{zxy}$, and $\chi^{(2)}_{zyx}$ as non-zero elements, *c*-type axial tensor has $\chi^{(2)}_{yyy}$, $\chi^{(2)}_{yxx}$, $\chi^{(2)}_{xyx}$, $\chi^{(2)}_{xxy}$, $\chi^{(2)}_{yzz}$, $\chi^{(2)}_{zyz}$, and $\chi^{(2)}_{zzy}$ as non-zero elements. In backscattering geometry, ignoring all the elements with *z* component, only *c*-type axial tensor of *mmm* point group has nonzero elements. As a result, MD SHG can probe magnetic transition of CrSBr monolayer. Observing that the magnetic field of electromagnetic wave propagating in free space is perpendicular to the electric field, parallel and cross polarized components of $\chi^{(2,c),eem}$ is written as

$$\begin{pmatrix}\chi_\parallel^{(2,c),eem}\\ \chi_\perp^{(2,c),eem}\end{pmatrix}$$

$$=\begin{pmatrix}\cos\theta & \sin\theta\\ -\sin\theta & \cos\theta\end{pmatrix}\begin{pmatrix}0 & 0 & \chi_{xxy}^{(2,c),eem} & \chi_{xyx}^{(2,c),eem}\\ \chi_{yxx}^{(2,c),eem} & \chi_{yyy}^{(2,c),eem} & 0 & 0\end{pmatrix}\begin{pmatrix}\cos\theta\cos(\theta+\pi/2)\\ \sin\theta\sin(\theta+\pi/2)\\ \cos\theta\sin(\theta+\pi/2)\\ \sin\theta\cos(\theta+\pi/2)\end{pmatrix}$$

$$=\begin{pmatrix}\cos\theta & \sin\theta\\ -\sin\theta & \cos\theta\end{pmatrix}\begin{pmatrix}0 & 0 & \chi_{xxy}^{(2,c),eem} & \chi_{xyx}^{(2,c),eem}\\ \chi_{yxx}^{(2,c),eem} & \chi_{yyy}^{(2,c),eem} & 0 & 0\end{pmatrix}\begin{pmatrix}-\cos\theta\sin\theta\\ \cos\theta\sin\theta\\ \cos^2\theta\\ -\sin^2\theta\end{pmatrix}$$

$$=\begin{pmatrix}\chi_{xxy}^{(2,c),eem}\cos^3\theta+\left(-\chi_{yxx}^{(2,c),eem}+\chi_{yyy}^{(2,c),eem}-\chi_{xyx}^{(2,c),eem}\right)\cos\theta\sin^2\theta\\ \left(-\chi_{yxx}^{(2,c),eem}+\chi_{yyy}^{(2,c),eem}-\chi_{xxy}^{(2,c),eem}\right)\cos^2\theta\sin\theta+\chi_{xyx}^{(2,c),eem}\sin^3\theta\end{pmatrix}$$

(S10)

which has the same functional form as $\chi^{(2),mee}$ of *mm2* point group.

For $\chi^{(2,c),mee}$,

$$\begin{pmatrix}\chi_\parallel^{(2,c),mee}\\ \chi_\perp^{(2,c),mee}\end{pmatrix}$$

$$=\begin{pmatrix}\cos\theta & \sin\theta\\ -\sin\theta & \cos\theta\end{pmatrix}\begin{pmatrix}0 & 0 & \chi_{xyx}^{(2,c),mee}\\ \chi_{yxx}^{(2,c),mee} & \chi_{yyy}^{(2,c),mee} & 0\end{pmatrix}\begin{pmatrix}\cos^2\theta\\ \sin^2\theta\\ 2\cos\theta\sin\theta\end{pmatrix}$$  (S11)

$$=\begin{pmatrix}\left(\chi_{yxx}^{(2,c),mee}+2\chi_{xyx}^{(2,c),mee}\right)\cos^2\theta\sin\theta+\chi_{yyy}^{(2,c),mee}\sin^3\theta\\ \chi_{yxx}^{(2,c),mee}\cos^3\theta+\left(\chi_{yyy}^{(2,c),mee}-2\chi_{xyx}^{(2,c),mee}\right)\cos\theta\sin^2\theta\end{pmatrix}.$$

These terms govern induced nonlinear magnetization in the material, which translates to the source term by $S\propto\mu_0\left(\nabla\times\frac{\partial M}{\partial t}\right)$ and can be written as

$$\begin{pmatrix}S_\parallel^{mee}(2\omega)\\ S_\perp^{mee}(2\omega)\end{pmatrix}\propto\begin{pmatrix}M_\perp^{mee}(2\omega)\\ -M_\parallel^{mee}(2\omega)\end{pmatrix}\propto\begin{pmatrix}\chi_\perp^{(2,c),mee}\\ -\chi_\parallel^{(2,c),mee}\end{pmatrix}$$

$$\propto\begin{pmatrix}\chi_{yxx}^{(2,c),mee}\cos^3\theta+\left(\chi_{yyy}^{(2,c),mee}-2\chi_{xyx}^{(2,c),mee}\right)\cos\theta\sin^2\theta\\ -\left(\chi_{yxx}^{(2,c),mee}+2\chi_{xyx}^{(2,c),mee}\right)\cos^2\theta\sin\theta-\chi_{yyy}^{(2,c),mee}\sin^3\theta\end{pmatrix}$$  (S12)

which now also has the same functional form. The total magnetic dipole second order susceptibility can be written as the sum of $\chi^{(2,c),eem}$ and $\chi^{(2,c),mee}$:

$$\begin{pmatrix} \chi_\parallel^{(2,c),\text{MD}} \\ \chi_\perp^{(2,c),\text{MD}} \end{pmatrix} = \begin{pmatrix} \chi_\parallel^{(2,c),eem} + \chi_\perp^{(2,c),mee} \\ \chi_\perp^{(2,c),eem} - \chi_\parallel^{(2,c),mee} \end{pmatrix} = \begin{pmatrix} A\cos^3\theta + (B+C)\cos\theta\sin^2\theta \\ (-A+B)\cos^2\theta\sin\theta - C\sin^3\theta \end{pmatrix} \quad \text{(S13)}$$

where

$$\begin{aligned} A &= \chi_{xxy}^{(2,c),eem} + \chi_{yxx}^{(2,c),mee} \\ B &= -\chi_{yxx}^{(2,c),eem} + \chi_{yyy}^{(2,c),eem} - 2\chi_{xyx}^{(2,c),mee} \\ C &= -\chi_{xyx}^{(2,c),eem} + \chi_{yyy}^{(2,c),mee} \end{aligned} \quad \text{(S14)}$$

As the equations show, polarization resolved measurement alone cannot differentiate between $\chi^{(2),eem}$ and $\chi^{(2),mee}$. Estimating relative orders of $\chi^{(2),eem}$ and $\chi^{(2),mee}$ requires understanding sizes of electric dipole and magnetic dipole transition matrix elements at $\omega$ and $2\omega$.

To achieve amplitude of second harmonic output, the source term must be integrated over the material thickness as the fundamental propagates through the medium. Since the material thickness is much smaller than the wavelengths, we can ignore phase matching conditions and not consider phase differences between second harmonic waves generated at different thicknesses. With these in consideration, second harmonic field amplitude is given as (*7*)

$$E(2\omega) = \frac{1}{4}\frac{i2\omega d}{2n_{2\omega}c}\chi^{(2)}E(\omega)^2 \quad \text{(S15)}$$

where $d$ is thickness of the material, which is approximated to 0.8 nm × number of layers. Modifications must be made to reflect gaussian-shaped beam and ultrafast pulse width.(*8*) However, precisely determining all the relevant parameters on-sample is not very practical. Instead, we compare SHG from samples and a reference, such as α-quartz, to obtain absolute values of $\chi^{(2)}$ (*9*).

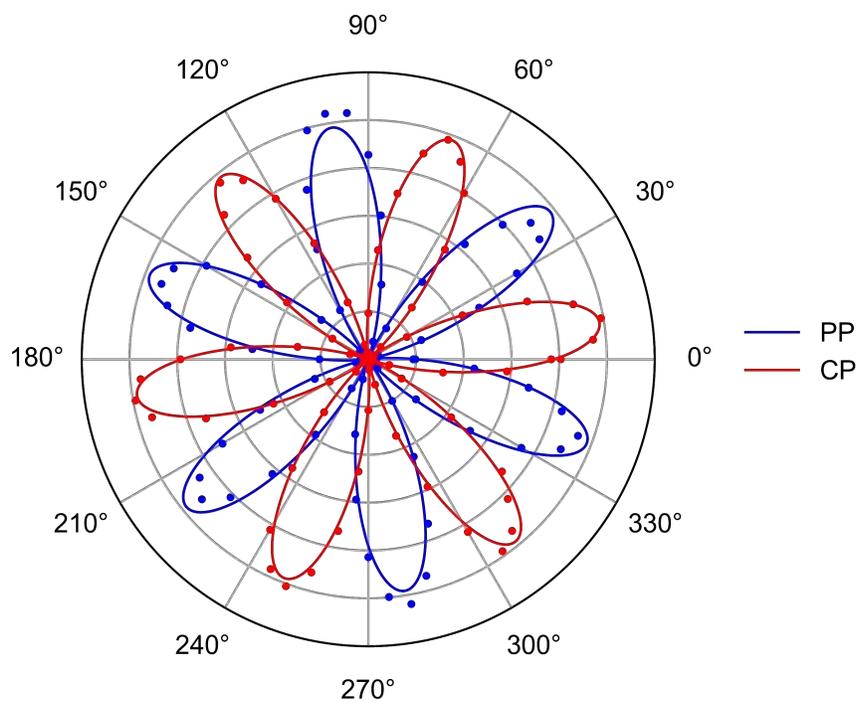

**Figure S9.** Polarization resolved SHG of MoS$_2$ monolayer.

**Temperature dependent SHG**

Second order susceptibility is proportional to the order parameters (*10, 11*) at arbitrary number of layers. We justify this by extending a generalized Ginzburg-Landau approach for SHG discussed by D. Sa, R. Valenti, and C. Gros (*12*). Time-averaged free energies corresponding to electric dipole and magnetic dipole components of SHG are given by (*13*)

$$\begin{aligned}
F^{eee} &= -\big[\chi_{ijk}^{eee}(2\omega,\omega,\omega)E_i^*(2\omega)E_j(\omega)E_k(\omega) \\
&\quad + \chi_{ijk}^{eee*}(2\omega,\omega,\omega)E_i(2\omega)E_j^*(\omega)E_k^*(\omega)\big] \\
F^{eem} &= -\big[\chi_{ijk}^{eem}(2\omega,\omega,\omega)E_i^*(2\omega)E_j(\omega)H_k(\omega) \\
&\quad + \chi_{ijk}^{eem*}(2\omega,\omega,\omega)E_i(2\omega)E_j^*(\omega)H_k^*(\omega)\big] \\
F^{mee} &= -\big[\chi_{ijk}^{eee}(2\omega,\omega,\omega)H_i^*(2\omega)E_j(\omega)E_k(\omega) \\
&\quad + \chi_{ijk}^{mee*}(2\omega,\omega,\omega)H_i(2\omega)E_j^*(\omega)E_k^*(\omega)\big]
\end{aligned} \quad (S16)$$

where $i, j, k$ are indices of the Cartesian coordinates.

In the Ginzberg-Landau formulation, the free energy must comply to the crystallographic symmetry of the material. To obtain expression for SHG that only exists in magnetically ordered states we write higher order SHG terms in the lowest order of order parameter. In FM monolayer, or a magnetically ordered odd number of layers, order parameter is $M$ which is *c*-type axial tensor of rank 1. The *eem* term of the free energy is described as

$$\begin{aligned}
F^{eem} = -[&\chi_{ijkl}^{cryst,(i)}(2\omega,\omega,\omega,0)E_i^*(2\omega)E_j(\omega)H_k(\omega) \\
+&\chi_{ijkl}^{cryst,(i)*}(2\omega,\omega,\omega,0)E_i(2\omega)E_j^*(\omega)H_k^*(\omega)]M_l
\end{aligned} \quad (S17)$$

where $\chi_{ijkl}^{cryst,(i)}$ is an *i*-type axial tensor of rank 4, for the free energy is an *i*-type scalar. Comparing it to the previous equation gives

$$\chi_{ijk}^{eem,(c)}(2\omega,\omega,\omega) = \chi_{ijkl}^{cryst,(i)}(2\omega,\omega,\omega,0)M_l \quad (S18)$$

which describes linear proportionality of $\chi_{ijk}^{eem,(c)}$ to $M$. The *mee* term follows similar derivation. In AFM bilayer, or a magnetically ordered even number of layers, order parameter is $T$ which is *c*-type polar tensor of rank 1. The *eee* term of the free energy can be written as

$$F^{eee} = -[\chi_{ijkl}^{cryst,(i)}(2\omega,\omega,\omega,0)E_i^*(2\omega)E_j(\omega)E_k(\omega)$$
$$+\chi_{ijkl}^{cryst,(i)*}(2\omega,\omega,\omega,0)E_i(2\omega)E_j^*(\omega)E_k^*(\omega)]T_l \quad (S19)$$

where $\chi_{ijkl}^{cryst,(i)}$ is an *i*-type polar tensor of rank 4. Comparing it to the previous equation gives

$$\chi_{ijk}^{eee,(c)}(2\omega,\omega,\omega) = \chi_{ijkl}^{cryst,(i)}(2\omega,\omega,\omega,0)T_l \quad (S20)$$

which describes linear proportionality of $\chi_{ijk}^{eee}$ to $T$. As a result, for both FM monolayer and AFM bilayer, or for any number of layer flake with magnetic ordering, the second order susceptibility is proportional to the order parameter.

Now we check the above interpretation agrees with second order susceptibility tensors from magnetic point groups. Point group *mmm* has polar tensor of rank 4 with 21 nonzero elements: *xxxx, yyyy, zzzz, xxyy* (and its 5 other permutations), *yyzz* (and its 5 other permutations), and *zzxx* (and its 5 other permutations). In the odd number of layers, order parameter $M$ is in $y$ direction. As a result, $\chi_{ijk}^{eem,(c)}$ inherits $l = y$ elements of $\chi_{ijkl}^{cryst,(i)}$, which gives *yyy, xxy, xyx, yxx, zzy, zyz*, and *yzz*. In the even number of layers, magnetic toroidal order parameter $T$ is in $x$ direction. Hence, $\chi_{ijk}^{eee,(c)}$ inherits $l = x$ elements of $\chi_{ijkl}^{cryst,(i)}$, which gives *xxx, xyy, yxy, xyy, xzz, zxz*, and *xzz*. These results agree with *c*-type tensors given in Table S2.

Lastly, in CrSBr, intralayer FM coupling is much stronger than interlayer AFM coupling, and we could approximate that magnetization in isolated monolayer and single layer in bulk show similar temperature dependence. As a result, $\chi^{(2)} \propto (1 - T/T_C)^\beta$ regardless of sample thickness. In the presence of magnetostrictive effect, $\chi^{(2)}$ is proportional to square of the order parameter (*14*). We do not observe such effect in our measurements.

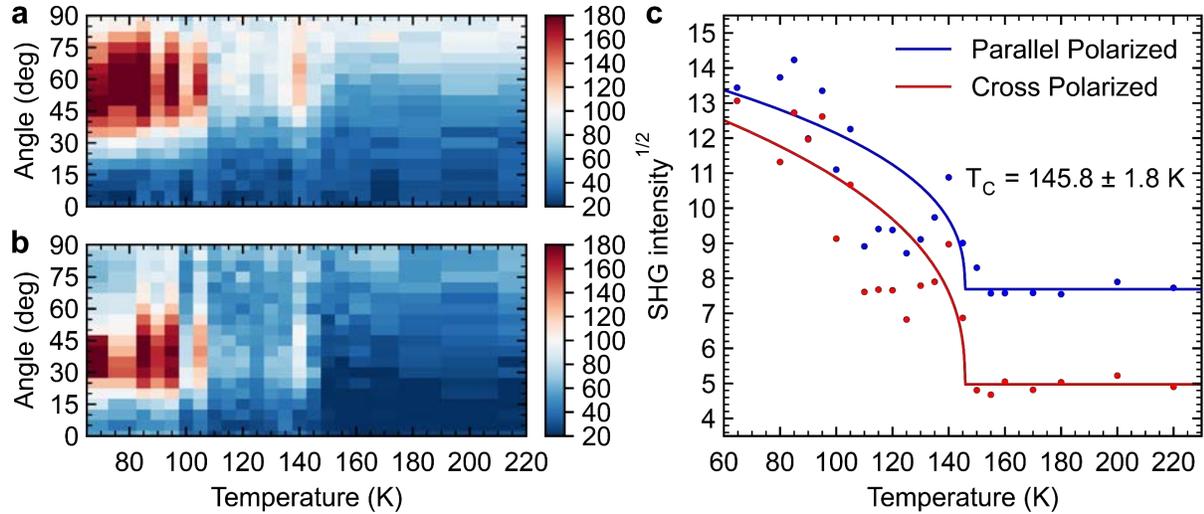

**Figure S10. (a)** Parallel-polarized and **(b)** cross-polarized temperature-dependent SHG for a CrSBr monolayer. Only 0° to 90° range is measured. **(c)** Square root of SHG intensity as a function of temperature. Fitting with $\chi^{(2)} \propto (1 - T/T_C)^\beta$ gives 146.5 K Curie temperature.

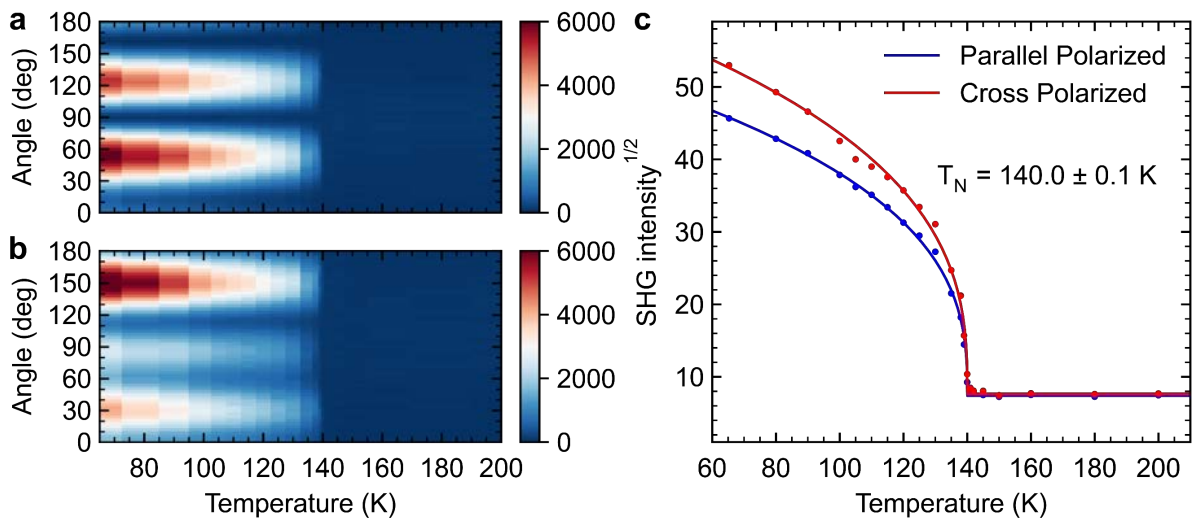

**Figure S11. (a)** Parallel-polarized and **(b)** cross-polarized temperature-dependent SHG for a CrSBr bilayer. **(c)** Square root of SHG intensity as a function of temperature. Fitting with $\chi^{(2)} \propto (1 - T/T_C)^\beta$ gives 140.3 K Néel temperature.

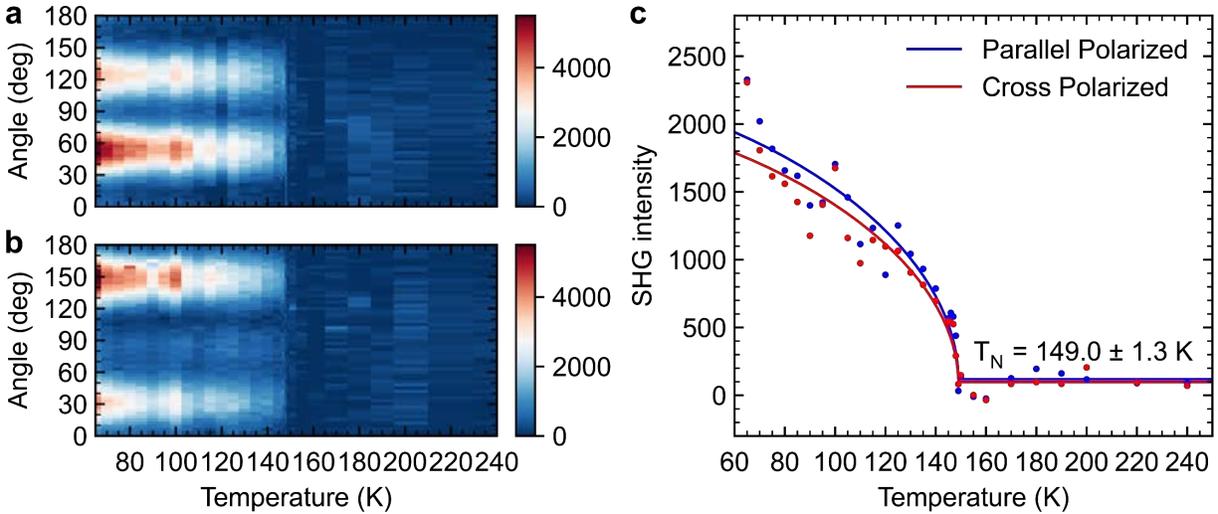

**Figure S12. (a)** Parallel-polarized and **(b)** cross-polarized temperature-dependent SHG for a different CrSBr bilayer sample. **(c)** Square root of SHG intensity as a function of temperature. Fitting with $\chi^{(2)} \propto (1 - T/T_c)^\beta$ gives 149.0 K Néel temperature.

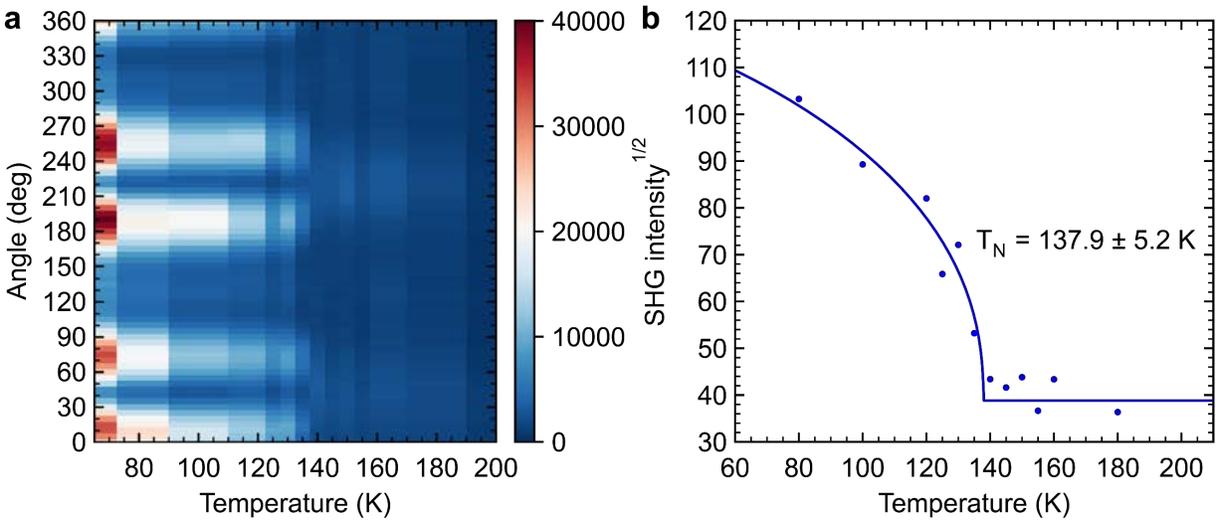

**Figure S13. (a)** Cross-polarized temperature-dependent SHG for a CrSBr 6 layer sample. **(b)** Square root of SHG intensity as a function of temperature. Fitting with $\chi^{(2)} \propto (1 - T/T_c)^\beta$ gives 137.9 K Néel temperature.

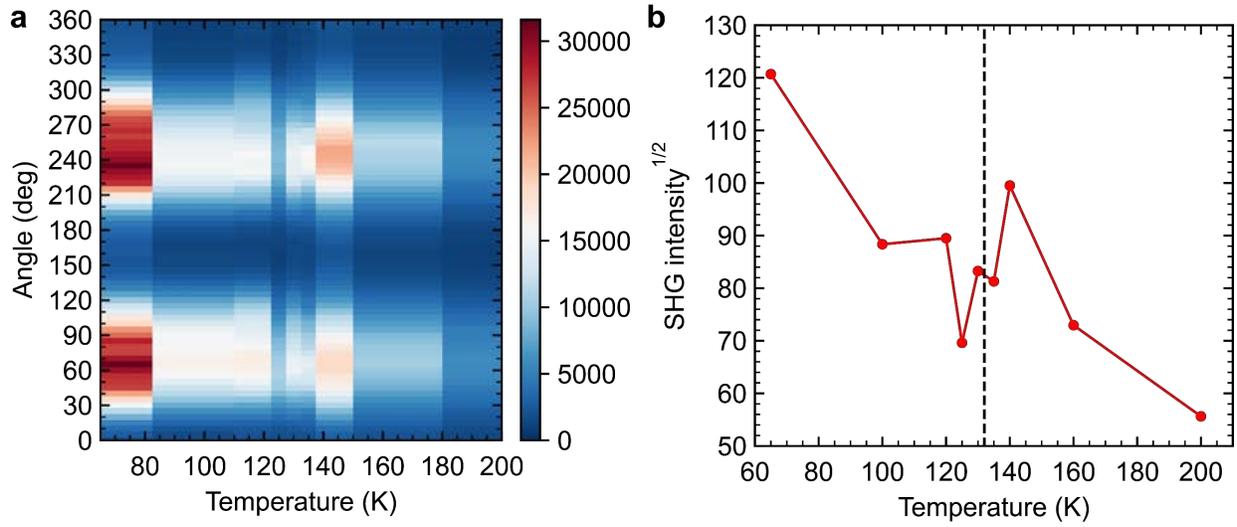

**Figure S14. (a)** Cross-polarized temperature-dependent SHG for a CrSBr thin bulk sample. **(b)** Square root of SHG intensity as a function of temperature gas a derivative shape across $T_N$ = 132 K.

**Interlayer geometry and magnetic coupling**

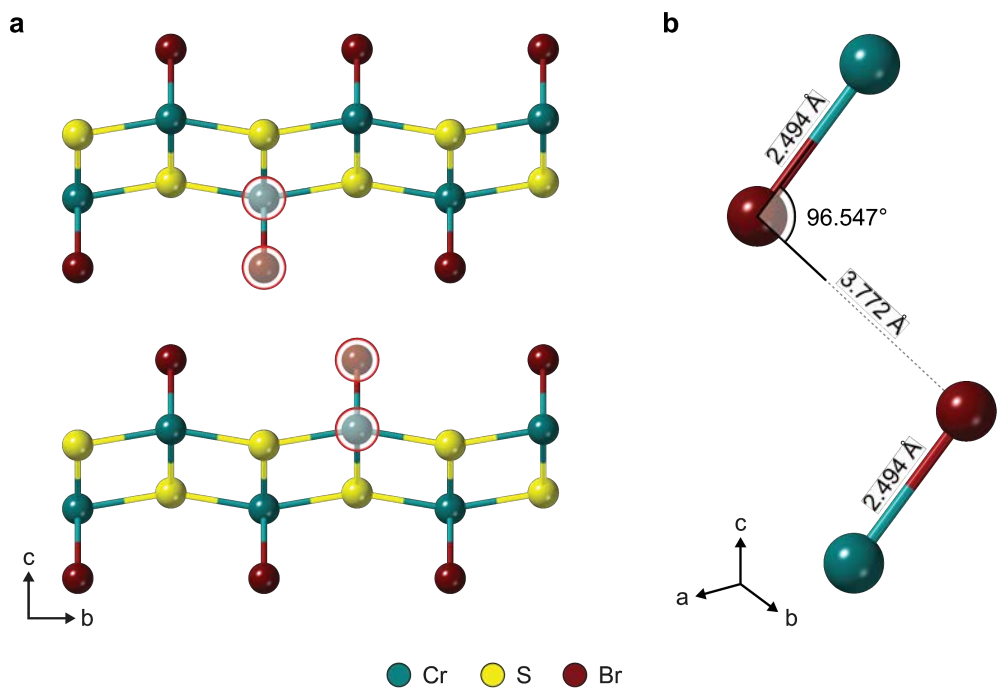

**Figure S15. (a)** CrSBr bilayer structure viewed along the crystallographic *a*-axis. The interlayer magnetic interaction is primarily antiferromagnetic super-superexchange (SSE) of Cr-Br···Br-Cr, exemplified by the atoms enclosed within red circles. **(b)** The four circled atoms from **(a)** viewed from a direction perpendicular to the plane containing all four atoms. These four atoms are the only symmetrically unique combination that have geometric configuration and close enough distances to host SSE in the intrinsic stacking. Cr-Br-Br angle is 96.547°, suggesting antiferromagnetic interlayer interaction.

**Heat capacity**

The measured heat capacity for magnetic materials contains contributions from lattice vibrations/rotations as well as local (dis)order induced by magnetic structure changes. For this reason, heat capacity measurements at temperatures near the magnetic ordering temperature of a material can be a valuable tool towards investigating the mechanism of magnetic ordering events, especially in systems that are highly magnetically anisotropic. The temperature- and magnetic field-dependent adiabatic molar heat capacity of CrSBr were recorded using a Quantum Design Physical Property Measurement System (PPMS) VersaLab between 100 and 200 K. The temperature-dependent heat capacity traces for CrSBr show two main features, a sharp peak at 132 K (zero field curve) and a broad feature around 160 K (zero field curve), both of which show magnetic field dependence. At increasing field strengths, the 132 K peak shifts to lower temperature and flattens into the baseline, while the broad 160 K feature shifts to higher temperature and sharpens. The peak at 132 K is attributed to the antiferromagnetic ordering event given the similarity in temperature between this peak and the reported AFM ordering temperature for CrSBr ($T_N$ = 132 K). However, the disappearance of this feature at magnetic fields above the reported saturation field for CrSBr ($H_C$ ~0.2 T at 120 K) suggests local ferromagnetic ordering takes place at temperatures above the $T_N$. It has previously been reported that low-dimensional systems show broad features in temperature-dependent heat capacity traces at temperatures above distinct magnetic ordering events due to short-range order induced by local magnetic coupling (*15*). For this reason, we assign the 160 K feature to the onset of local ferromagnetic ordering, likely within a single or few layer(s) of CrSBr; at applied fields less than $H_C$, the ferromagnetic domains (layers) couple antiferromagnetically leading to the heat capacity peak ~134 K and at applied fields greater than $H_C$, the applied field aligns individual domains (layers) ferromagnetically such that the AFM coupling event is absent.

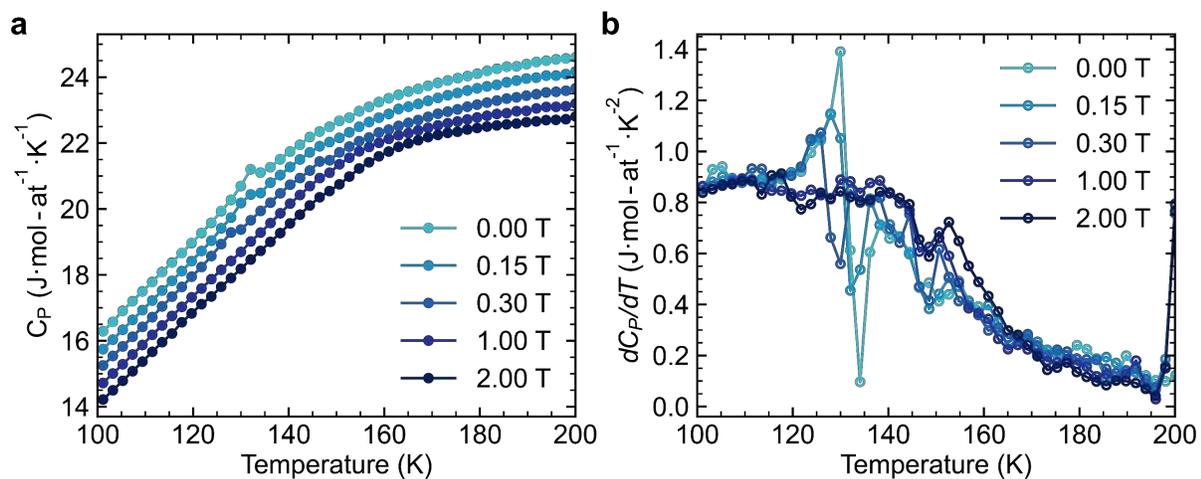

**Figure S16.** (**a**) Specific heat capacity of CrSBr at various external magnetic fields. Curves are vertically offset by 0.5 for clarity. At 0 T, a sharp peak at 132 K correspond to the bulk Néel temperature. (**b**) First derivative of heat capacity to show onset ~160 K indicating local magnetic transition.

# References

1. J. Beck, Über Chalkogenidhalogenide des Chroms Synthese, Kristallstruktur und Magnetismus von Chromsulfidbromid, CrSBr. *Z. Anorg. Allg. Chem.* **585**, 157 (1990).
2. R. C. Clark, J. S. Reid, The analytical calculation of absorption in multifaceted crystals. *Acta Crystallogr., Sect. A: Found. Crystallogr.* **51**, 887 (1995).
3. Rigaku Oxford Diffraction, CrysAlisPro Version 1.171.38.46 (2015).
4. G. M. Sheldrick, Crystal structure refinement with SHELXL. *Acta Crystallogr., Sect. C: Struct. Chem.* **71**, 3 (2015).
5. O. V. Dolomanov, L. J. Bourhis, R. J. Gildea, J. A. K. Howard, H. Puschmann, OLEX2: a complete structure solution, refinement and analysis program. *J. Appl. Crystallogr.* **42**, 339 (2009).
6. O. A. Ajayi, J. V. Ardelean, G. D. Shepard, J. Wang, A. Antony, T. Taniguchi, K. Watanabe, T. F. Heinz, S. Strauf, X. Y. Zhu, J. C. Hone, Approaching the intrinsic photoluminescence linewidth in transition metal dichalcogenide monolayers. *2D Mater.* **4**, 031011 (2017).
7. N. Kumar, S. Najmaei, Q. N. Cui, F. Ceballos, P. M. Ajayan, J. Lou, H. Zhao, Second harmonic microscopy of monolayer MoS2. *Phys. Rev. B* **87**, 161403(R) (2013).
8. Y.-R. Shen, *The principles of nonlinear optics*. (Wiley-Interscience, New York, 1984).
9. Y. Li, Y. Rao, K. F. Mak, Y. You, S. Wang, C. R. Dean, T. F. Heinz, Probing symmetry properties of few-layer MoS2 and h-BN by optical second-harmonic generation. *Nano Lett.* **13**, 3329 (2013).
10. M. Fiebig, D. Frohlich, B. B. Krichevtsov, R. V. Pisarev, Second harmonic generation and magnetic-dipole-electric-dipole interference in antiferromagnetic Cr2O3. *Phys. Rev. Lett.* **73**, 2127 (1994).
11. M. Matsubara, C. Becher, A. Schmehl, J. Mannhart, D. G. Schlom, M. Fiebig, Optical second- and third-harmonic generation on the ferromagnetic semiconductor europium oxide. *J. Appl. Phys.* **109**, 07C309 (2011).
12. D. Sa, R. Valentí, C. Gros, A generalized Ginzburg-Landau approach to second harmonic generation. *Eur. Phys. J. B* **14**, 301 (2000).
13. P. S. Pershan, Nonlinear Optical Properties of Solids: Energy Considerations. *Phys. Rev.* **130**, 919 (1963).
14. M. Fiebig, D. Frohlich, T. Lottermoser, V. V. Pavlov, R. V. Pisarev, H. J. Weber, Second harmonic generation in the centrosymmetric antiferromagnet NiO. *Phys. Rev. Lett.* **87**, 137202 (2001).
15. M. A. McGuire, G. Clark, S. Kc, W. M. Chance, G. E. Jellison, V. R. Cooper, X. Xu, B. C. Sales, Magnetic behavior and spin-lattice coupling in cleavable van der Waals layered CrCl3 crystals. *Phys. Rev. Mater.* **1**, 014001 (2017).